\documentclass{elsarticle}

\usepackage{amsmath,amsfonts,amssymb,bm}
\usepackage{graphicx}
\usepackage{color}
\usepackage{algorithm}
\usepackage{caption}
\usepackage{subcaption}
\usepackage{ulem}
\usepackage{algpseudocode}
\usepackage{booktabs}
\usepackage{tabularx}
\usepackage{colortbl}
\usepackage{hhline}
\usepackage{epstopdf}
\usepackage{xcolor}

\journal{Journal of Sound and Vibration}

\begin{document}
	
	\begin{frontmatter}
		
		\title{ABC for model selection and parameter estimation of drill-string bit-rock interaction models and stochastic stability}
		
		\author[UFRJ]{D.A. Castello \corref{cor1}}
		\ead{castello@mecanica.coppe.ufrj.br}
		

		\author[UFRJ,Bristol]{T.G. Ritto}

		\cortext[cor1]{Corresponding author}
		
		\address[UFRJ]{Department of Mechanical Engineering - Universidade Federal do Rio de Janeiro, Rio de Janeiro, Brazil}
		
		\address[Bristol]{Department of Mechanical Engineering - University of Bristol, Bristol, UK}
		

		
		

		\begin{abstract}
			The bit-rock interaction considerably affects the dynamics of a drill string. One critical condition is the stick-slip oscillations, where torsional vibrations are high; the bit angular speed varies from zero to about two times (or more) the top drive nominal angular speed. In addition, uncertainties should be taken into account when calibrating (identifying) the bit-rock interaction parameters. This paper proposes a procedure to estimate the parameters of four bit-rock interaction models, one of which is new, and at the same time select the most suitable model, given the available field data. The approximate Bayesian computation (ABC) is used for this purpose. An approximate posterior probability density function is obtained for the parameters of each model, which allows uncertainty to be analyzed. Furthermore, the impact of the uncertainties of the selected models on the torsional stability map (varying the nominal top drive angular speed and the weight on bit) of the system is evaluated.
			\end{abstract}
		
		
	\end{frontmatter}
	

\section{Introduction}

Drilling is an important stage in oil and gas exploration. To make drilling happen a long and slender drill-string rotates and perforates different rock formations. Many different drill-string dynamic models can be found in the literature. Some authors use a pure torsional model \cite{Kreuzer2012,Vromen2017,Monteiro2017,Ritto2017,Tian2019}, while others consider coupling mechanisms. For instance, lateral-torsional \cite{Abbassian1998,Yigit1998,Volpi2021}, axial-torsional \cite{Richard2007,Lobo2020}, and fully lateral-axial-torsional coupled vibrations \cite{Tucker1999,Khulief2005,Ritto2009,deMoraes2019}. One can find single torsional models, lumped parameters representation, and continuous systems discretized using the finite element method. For the purposes of the present investigation, we will consider a single degree of freedom torsional model that suffices to represent the phenomenon under analysis \cite{Ritto2018}. It should be highlighted that the following proposed methodology is not limited to a simple lumped parameter model; it can (and should) be applied to a fully coupled system with different bit-rock interaction mechanisms. \ref{sec:appendix} presents a torsional finite element model confirming that a one degree-of-freedom model is enough for the cases analyzed here.

Concerning the bit-rock interaction, we can mention the coupled axial-torsional model developed by Detournay and Defourny \cite{Detournay1992} and further developed by Richard \textit{et al.} \cite{Richard2007}, which takes into account the cutting and friction mechanisms, and the heuristic model proposed by Tucker and Wang \cite{Tucker2003}. For the pure torsional vibration, which is the case analyzed here, we will consider three models found in the literature: Tucker and Wang \cite{Tucker97}, Navarro-Lopez and Suarez \cite{Navarro2004} and Ritto \textit{ et al.} \cite{Ritto2017}. In addition, we propose a new model that combines elements of the models found in \cite{Tucker97,Navarro2004}.

The bit rock-interaction model should be identified with field data \cite{Ritto2017,Real2018} or lab experiments \cite{Kapitaniak2016,Lobo2020}. If uncertainties are considered, the maximum likelihood \cite{Aldrich1997,Ritto2010} or the Bayesian approach \cite{Kaipio2006,Ritto2015b} might be employed, for example.

The contribution of the present paper is to apply a Bayesian approach for estimating the parameters and selecting the most adapted bit-rock interaction model in a torsional dynamic drill string problem, considering uncertainties and analyzing (stochastic) stability. Field data obtained from a 5km drill-string is used in this endeavor \cite{Ritto2017}. Four models with different associated parameters are analyzed: model 1 uses hyperbolic tangent \cite{Tucker97}, model 2 uses the exponential function \cite{Navarro2004}, model 3 combines hyperbolic tangent with the exponential function (new model), and model 4 uses a cubic polynomial \cite{Ritto2017}.

The remainder of the paper is organized as follows. Section \ref{sec:models} presents the dynamical system and the nonlinear bit-rock interaction models analyzed. Section \ref{sec:stability} briefly introduces the stability analysis framework. Section \ref{sec:ABC} depicts the Bayesian strategy employed for parameter estimation and model selection. The results are shown in Section \ref{sec:results} and the conclusions are made in the last section. Finally, \ref{sec:appendix} shows a finite element (FE) formulation for the torsional drill-string and draws some stability maps.

\label{sec:models}

\section{Bit-rock interaction models}
\label{sec:models}

\begin{figure}[!htb]
	\centering
	\includegraphics[height=5.8cm]{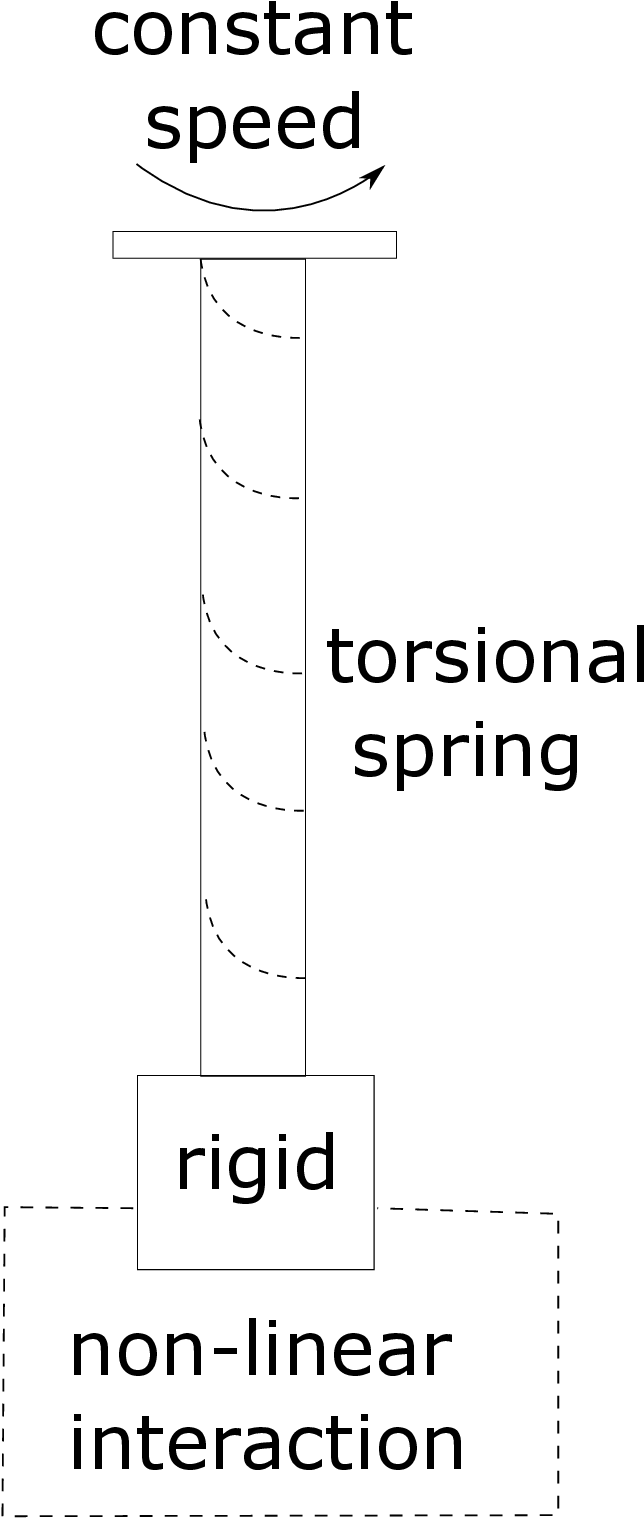}
	\caption{The basic components of a drill-string model: rotary table at the top, elastic column along the core, rigid Bottom Hole Assembly (BHA) at the bottom. The mechanical interactions between the soil and the drill-bit are generally described by nonlinear phenomenological models.}
	\label{drillstring}
\end{figure}

Bit-Rock interaction models play an important role when simulating drill-string dynamics, especially in torsional stability. Figure \ref{drillstring} presents the basic components of a drill-string model, namely: a rotary table at the top, an elastic column in the core, and a rigid element at the bottom describing a set of components which is generally named as Bottom Hole Assembly (BHA). Assuming the drill-string dynamics may be properly described by torsion effects only as in \cite{Ritto2018}, one may write its governing equation based on a lumped-parameter description as follows:     
\begin{equation}
\label{eq:governing}
I_{eq}\ddot{\theta}(t) + c_{eq}\dot{\theta}(t) + k_{eq}{\theta}(t) = k_{eq} \, \Omega t +c_{eq} \, \Omega - T_{bit}(\dot{\theta}(t))
\end{equation}
\noindent where $I_{eq}$, $c_{eq}$ and $k_{eq}$ are the moment of inertia, viscous damping coefficient and torsional stiffness of the equivalent lumped-parameter model. Regarding $\Omega$ and $\theta(t)$, they are the constant angular speed of the rotary table and the angle of the bit, respectively. Torque $T_{bit}(\dot{\theta}(t))$ in Eq.\ref{eq:governing} is used here to represent nonlinear dynamic effects associated with the interaction of the soil with the bit during drilling operations. A FE model for the torsional system can be found in
\ref{sec:appendix}.

Regarding the description of the interactions between the soil and the bit, a set of 
four candidate models $\{\mathcal{M}_1, \ldots, \mathcal{M}_4\}$ are used in this work to describe such phenomena. These models are defined by specific  relations between the torque on bit $T_{bit}(t)$ and  the bit speed $\dot{\theta}(t)$ in Eq.\ref{eq:governing}. The set of models are introduced next   (assuming  $\dot{\theta}(t)>0$):

The first model was proposed by Tucker and Wang \cite{Tucker97}, and slightly modified by Nogueira and Ritto \cite{Nogueira2018}. It is described as follows

\begin{equation}
T_{bit}^{(1)}(\dot{\theta}) =r b_0\left( \tanh(b_1\dot{\theta}) + \frac{b_2\dot{\theta}}{(1+b_3\dot{\theta}^2)}\right)\,, 
\end{equation}
\noindent where the ratio $r=\mathcal{W}/\mathcal{W}_{ref}$ is given by the weight on bit $\mathcal{W}$, which is the axial force applied at the bit (assumed constant), divided by the reference value $\mathcal{W}_{ref}$; the field data used for calibration is observed for $\mathcal{W}=\mathcal{W}_{ref}$, hence $r=1$ \cite{Ritto2017}. The superscript $(\bullet)^{(k)}$ in  $T_{bit}^{(k)}$  refers to the $k$-th model, namely $\mathcal{M}_k$. The first model is characterized by four parameters $\boldsymbol{\phi} = \{b_0,b_1,b_2,b_3\}^{\rm T}$ which should be properly estimated through inverse analysis \cite{Kaipio2006}. One may identify $rb_0$ as the limit for $T_{bit}^{(1)}$ when considering $\dot{\theta}(t)$ tending to infinity. 

%

The second model $\mathcal{M}_2$ was proposed by Navarro-Lopez and Suarez \cite{Navarro2004} and it is described as follows
\begin{equation}
T_{bit}^{(2)}(\dot{\theta}) =r((T_{sb}-T_{cb})\exp(-G_b\dot{\theta})+T_{cb})\,,
\label{eq:M2}
\end{equation}
\noindent This model is characterized by three parameters, namely $\boldsymbol{\phi} = \{T_{cb}, T_{sb}, G_b  \}^{\rm T}$. In particular, this model allows one to identify its maximum static torque as $r T_{sb}$ and its maximum dynamic torque as $rT_{cb}$; moreover, as $\dot{\theta}(t)$ tends to infinity, one may also characterize the decaying profile of the torque as a function of $G_b$. 

\begin{figure}[H]
\begin{center}
\includegraphics[scale=.45]{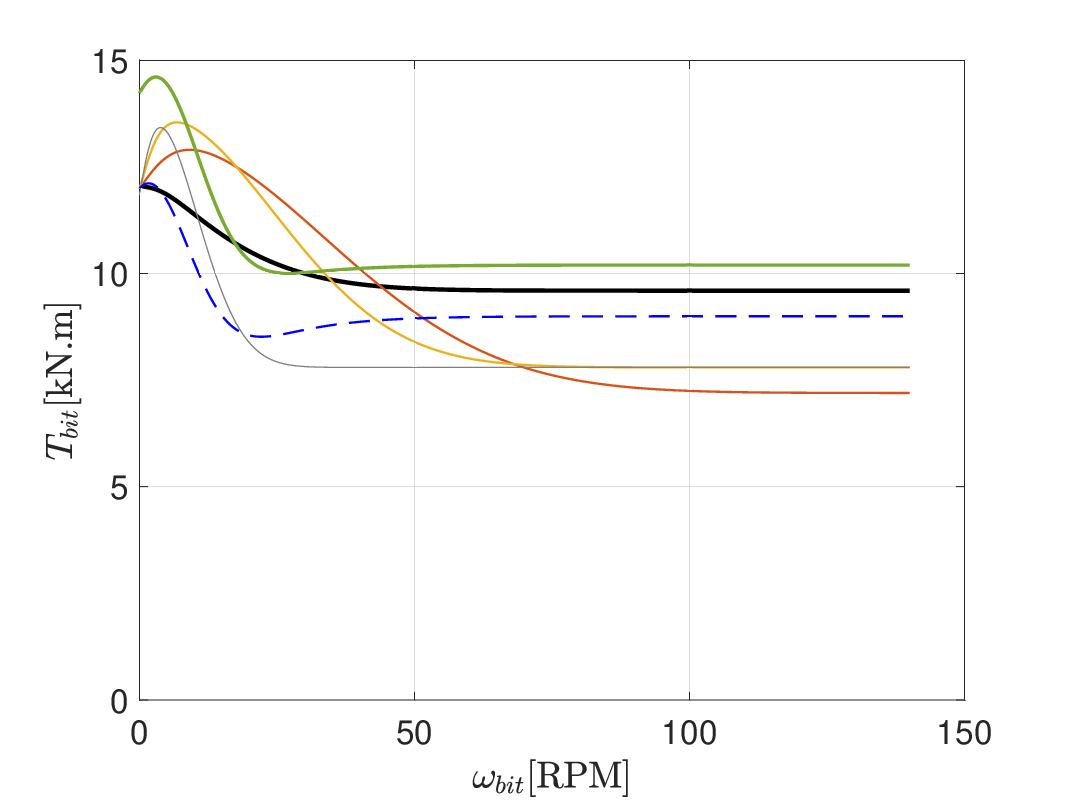}
\caption{Some torque profiles provided by model $\mathcal{M}_3$ obtained with six random samples of $\boldsymbol{\phi} = \{a_0,\ldots,a_5\}^{\rm T}$ and $r=1$.}
\label{fig:samples_M3}
\end{center}
\end{figure}

The third model $\mathcal{M}_3$ to be considered is new. {It combines three functions, namely: an exponential function, a constant gain and a hyperbolic tangent function as shown in Eq.\ref{eq:M3} }
\begin{equation}
T_{bit}^{(3)}(\dot{\theta})=r(a_0\exp(-a_1(\dot{\theta}-a_2)^2)+a_3-a_4\tanh(a_5\dot{\theta}))\,.
\label{eq:M3}
\end{equation}
\noindent Model $\mathcal{M}_3$ provides a long-term dynamic torque that is defined as a function of the constant gain $a_3$ and the amplitude of the hyperbolic function $-a_4$, hence $r(a_3 - a_4)$. As for the torque behavior around the static limit, it is governed by the static gain $a_3$, the amplitude $a_0$ and smoothness parameter $a_1$ that characterizes the exponential function.  More specifically, different patterns may be obtained around the static limit such as a fully-decreasing function after $\dot{\theta}$=0 or even a bump-like pattern followed by a decreasing function, for example. In this regard, Fig.(\ref{fig:samples_M3}) presents some torque profiles provided by model $\mathcal{M}_3$ which are obtained with different random samples of the parameter vector $\boldsymbol{\phi} = \{a_0,\ldots,a_5\}^{\rm T}$, which should be properly estimated by inverse analysis. 

The fourth model $\mathcal{M}_4$ is inspired by the model proposed by Ritto and Aguiar \cite{Ritto2017} and is a cubic polynomial as shown next 
\begin{equation}
T_{bit}^{(4)}(\dot{\theta})=r(c_0+c_1\dot{\theta}+c_2\dot{\theta}^2+c_3\dot{\theta}^3)\,,
\end{equation}
\noindent where $\boldsymbol{\phi} = \{c_0,c_1,c_2,c_3\}^{\rm T}$ are model parameters that should be identified by inverse analysis. In particular, this model provides $r c_0$ as its static torque  and it does not present an upper/lower limit for the dynamic torque.


\section{Stability analysis}\label{sec:stability}


The stability of the system should be analyzed for different operational parameter values ($\Omega,\mathcal{W}$) where $\Omega$ is the angular speed of the rotary table and $\mathcal{W}$ is the weight on bit. {\textcolor{black}{For this purpose, firstly one rewrites Eq.\ref{eq:governing} in the state space, with state vector $\textbf{x}(t)=(\theta(t),\dot{\theta}(t))$, as follows }} 
\begin{equation}
    \vspace{.1cm}\dot{\textbf{x}}(t)= \textbf{f}(\textbf{x}(t))=
    \left(\begin{array}{c}
         \dot{\theta}(t)  \\
          \frac{k_{eq}}{I_{eq}}(\Omega t - {\theta}(t)) +  \frac{c_{eq}}{I_{eq}}(\Omega  - \dot{\theta}(t))- (\frac{1}{I_{eq}})T_{bit}(\dot{\theta}(t))
    \end{array}\right)\,,
\end{equation}
{\textcolor{black}{Secondly, we consider the system operating at the point of dynamic equilibrium $\textbf{x}^*=( \Omega t- \theta_0,\Omega)$ such that $\mathbf{f}(\mathbf{x}^*) = (\Omega, 0 )$. This equilibrium point leads to $T_{bit}(\Omega) = k_{eq}\theta_0$ meaning that the torque on bit balances the elastic forces; further, the column rotates without torsional oscillations. Finally, aiming at analyzing the system stability around the dynamic equilibrium, we linearize the governing equations around $\mathbf{x}^*$ assuming that $\mathbf{x}(t) = \mathbf{x}^*(t) + \Delta \mathbf{x}(t)$ such that}}
\begin{equation}
\Delta\dot{\textbf{x}}(t)=\textbf{A}(\Omega,\mathcal{W}) \Delta\textbf{x}(t)\,,
\end{equation}
{\textcolor{black}{\noindent where the state matrix $\mathbf{A}$ (Jacobian matrix) is given as follows:}}
\begin{equation}
   \textbf{A}(\Omega,\mathcal{W})= \nabla_{\mathbf{x}} {\mathbf{f}}{\bigg{\vert}}_{\mathbf{x} = \mathbf{x}^*} = \left( \begin{array}{cc}
        0 & 1 \\
       -\omega_n^2  & -2\omega_n\xi - (\partial T_{bit}/\partial \dot{\theta})/I_{eq}
    \end{array}\right) {\bigg{\vert}}_{\mathbf{x} = \mathbf{x}^*}
\end{equation}
{\textcolor{black}{in which the natural frequency and the damping ratio are given by $\omega_n=\sqrt{k_{eq}/I_{eq}}$ and 
$\xi=c_{eq}/(2\sqrt{I_{eq}k_{eq}})$. If the real part of the eigenvalues of $\textbf{A}$ are all negative, the system is stable, otherwise it is unstable.  The notation $\textbf{A}(\Omega,\mathcal{W})$ highlights the fact that parametric stability analysis will be performed for a set of predefined operational parameters ($\Omega,\mathcal{W}$).}} {\textcolor{black}{It is  know that increasing $\mathcal{W}$ (i.e. increasing $r$) and decreasing $\Omega$ tend to destabilize the system \cite{Nogueira2018}.}}

The torque derivative for each one of the bit-rock interaction models are shown below (considering $\dot{\theta}>0$).


\begin{equation}
    \frac{\partial T_{bit}^{(1)}}{\partial\dot{\theta}}=-rb_0(b_1(\tanh(b_1\dot{\theta})^2 - 1) - b_2/(1+b_3\dot{\theta}^2) + (2b_2b_3\dot{\theta}^2)/(1 + b_3\dot{\theta}^2)^2)\,,
\end{equation}

\begin{equation}
    \frac{\partial T_{bit}^{(2)}}{\partial\dot{\theta}}= -r (T_{sb} - T_{cb})G_b \exp(-G_b \dot{\theta}) \,,
\end{equation}

\begin{equation}
    \frac{\partial T_{bit}^{(3)}}{\partial\dot{\theta}}=r(a_4 a_5(\tanh(a_5 \dot{\theta})^2 - 1) + (2 a_2 - 2 \dot{\theta}) a_0 a_1\exp{(-a_1(\dot{\theta}-a_2)^2)})\,,
\end{equation}

\begin{equation}
    \frac{\partial T_{bit}^{(4)}}{\partial\dot{\theta}}= r(c_1+2c_2\dot{\theta}+3c_3\dot{\theta}^2)\,.
\end{equation}

Finally, the procedure to verify the stability is similar if a FE model is considered; see 
\ref{sec:appendix}.

\section{Approximate Bayesian Computation (ABC)}\label{sec:ABC}

There is not much research that uses ABC for parameter estimation and model selection in the area of structural dynamics \cite{Chiachio2014,Abdessalem2018,Ritto2022}.  Assuming that a set of measured data $\mathbf{y} \in \mathbb{R}^d$ is available from a drill-string and considering the set of candidate models  $\mathbb{M} = \{\mathcal{M}_1, \mathcal{M}_2, \mathcal{M}_3, \mathcal{M}_4 \}$ shown in Section \ref{sec:models}, 
model learning is pursued using the Bayesian approach, which states that

\begin{equation}
\pi(\boldsymbol{\phi} \, | \textbf{y},\mathcal{M}_k ) = \frac{\pi( \textbf{y} \, |  \boldsymbol{\phi},\mathcal{M}_k) \, \pi_0(\boldsymbol{\phi}|\mathcal{M}_k)}{\pi({\textbf{y}|\mathcal{M}_k)}} \propto \pi( \textbf{y} \, |  \boldsymbol{\phi},\mathcal{M}_k) \, \pi_0(\boldsymbol{\phi}|\mathcal{M}_k)\,,
\label{eq:bayes}
\end{equation}
%
%
where $\boldsymbol{\phi} \in \mathbb{R}^n$ is the vector composed of the parameters of  model $\mathcal{M}_k$.  Regarding $\pi(\boldsymbol{\phi} \, | \textbf{y},\mathcal{M}_k )$, $\pi( \textbf{y} \, |  \boldsymbol{\phi},\mathcal{M}_k) $ and $\pi_0(\boldsymbol{\phi}|\mathcal{M}_k)$, they correspond to the posterior density, likelihood function and prior density  for model $\mathcal{M}_k$, respectively.

Let us adopt an additive observation model as shown next 
\begin{equation}
	\textbf{y} = \textbf{A}(\boldsymbol{\phi}) + \textbf{e}
	\label{eq:modelo-obs}\,,
\end{equation}

\noindent in which the operator $\textbf{A}: \boldsymbol{\phi} \in \mathbb{R}^n \mapsto \mathbb{R}^d$ provides model predictions and $\textbf{e} \in \mathbb{R}^d$ represents measurement and modeling errors. Further, assuming that $\boldsymbol{\phi}$ and $\mathbf{e}$ are mutually independent random variables, the likelihood function $\pi(\mathbf{y} | \boldsymbol{\phi})$ in Eq.(\ref{eq:bayes}) casts as follows
\cite{Kaipio2006}
\begin{equation}
\pi(\textbf{y} \, | \, \boldsymbol{\phi}) = \pi_{{e}}( \textbf{y} -\textbf{A}(\boldsymbol{\phi}))\,,
\label{eq:llkd}
\end{equation}

\noindent where $\pi_e(\mathbf{q})$ denotes the density of $\mathbf{e}$ evaluated at the argument $\mathbf{q}$. 
Unfortunately, as in many cases, here we do not know the structure of the error $\mathbf{e}$ or its density $\pi(\mathbf{e})$. Consequently, we cannot explicitly write the likelihood function $\pi(\textbf{y} \, | \, \boldsymbol{\phi})$ in Eq.(\ref{eq:llkd}). To circumvent this problem, we apply the Approximate Bayesian Computation (ABC) 	which was conceived aiming at inferring posterior densities where likelihood functions are computationally intractable or too costly to be evaluated \cite{Toni2008}.

In the ABC framework one replaces the calculation of the likelihood with a comparison between measured data $\mathbf{y}$ and model predictions $\mathbf{A}(\boldsymbol{\phi})$ based on a suitable distance function/metric $\rho: \mathbb{R}^d \times \mathbb{R}^d \mapsto \mathbb{R}$.  In this sense, the outputs of ABC algorithms correspond to a set of samples from a density $\pi(\boldsymbol{\phi} \, | \rho(\textbf{y}, \textbf{A}(\boldsymbol{\phi}) \leq \varepsilon)$ as follows
\begin{equation}
\pi(\boldsymbol{\phi} \, | \rho(\textbf{y}, \textbf{A}(\boldsymbol{\phi}) \leq \varepsilon) \propto \pi( \rho(\textbf{y}, \textbf{A}(\boldsymbol{\phi})) \leq \varepsilon)\times \pi_0(\boldsymbol{\phi})
\label{eq:bayes_ABC}
\end{equation}

 \noindent where $\varepsilon$ defines a tolerance/threshold acceptance level for a candidate sample $\boldsymbol{\phi}$.   Therefore, an ABC algorithm provides an {\it{approximation}} for the density $\pi(\boldsymbol{\phi} | \mathbf{y})$. Nevertheless, it should be emphasized that if the tolerance $\varepsilon$ in Eq.(\ref{eq:bayes_ABC}) is sufficiently small, then the density $\pi(\boldsymbol{\phi} \, | \rho(\textbf{y}, \textbf{A}(\boldsymbol{\phi})) \leq \varepsilon)$ will be a good approximation of the target posterior density $\pi(\boldsymbol{\phi} | \mathbf{y})$ in Eq.(\ref{eq:bayes}) \cite{Abdessalem2018,Toni2008, Abdessalem2019,ABCHandbook}.

Regarding the metric $\rho$ that could be used  in Eq.(\ref{eq:bayes_ABC}), as stated by  Toni {\it{et al.}} \cite{Toni2008}, these may vary according to the problem at hands and in principle they could consider either raw data comparisons or summary statistics comparisons, for example.  In the present work, we have time-domain data and we adopt a relative comparison of measured data $\mathbf{y}$ and model predictions computed at a candidate parameter vector $\boldsymbol{\phi}^*$ as a metric as shown next

\begin{equation}
\rho(\textbf{y}, \textbf{A}(\boldsymbol{\phi}^*))=\frac{||\textbf{y} -\textbf{A}(\boldsymbol{\phi}^*) ||^2_2}{||\textbf{y}  ||^2_2}\,.
\label{eq:metric}
\end{equation}

As for the tolerance $\varepsilon$ used in Eq.(\ref{eq:bayes_ABC}), one should set its value based on the intended level of agreement between measurements and predictions.  Further comments about this choice in the context of our work will be provided in Section \ref{sec:results}.

Concerning the posterior probability of model $\mathcal{M}_k$ given data $\textbf{y}$, namely $\mathbb{P}(\mathcal{M}_k)$, one may write \cite{Abdessalem2018,Beck2004,Ritto2015,Castello2018}

\begin{equation}
    \mathbb{P}(\mathcal{M}_k|\textbf{y})\propto \mathbb{P}_0(\mathcal{M}_k) \times \pi(\textbf{y}|\mathcal{M}_k) =  \mathbb{P}_0(\mathcal{M}_k)\times \int  \pi(\textbf{y}|\boldsymbol{\phi}, \mathcal{M}_k)  \pi_0(\boldsymbol{\phi}|\mathcal{M}_k)  d \boldsymbol{\phi}\,,
\end{equation}

\noindent where $\mathbb{P}_0(\mathcal{M}_k)$ is the prior model  {probability} and $\pi(\textbf{y}|\mathcal{M}_k)$ is called the evidence of  model $\mathcal{M}_k$.

\begin{algorithm}
\caption{ABC {\it{rejection sampler}}}
\begin{algorithmic}[1]

    \State Initialize $\varepsilon_1> \varepsilon_2 > \ldots, \varepsilon_G$ 
    \State Set the population indicator $g=0$
    \State Set the particle indicator $i=1$
    \State Sample $\mathcal{M}^*$ from $\mathbb{P}(\mathcal{M})$ (candidate model)
    \State Sample $\boldsymbol{\phi}^{*}$ from $\pi_0(\boldsymbol{\phi} | \mathcal{M}^*)$ (prior distribution)
    \State Simulate  $\mathbf{A}(\boldsymbol{\phi}^{*}|\mathcal{M}^*)$ (model prediction)

\If {$\rho(\mathbf{y}, \mathbf{A}(\boldsymbol{\phi}^{*}|\mathcal{M}^*)) \geq \varepsilon_g$} {Return to 4}
\Else 
\State {Set $\mathcal{M}_g^{(i)} = \mathcal{M}^*$ and add $\boldsymbol{\phi}^{*}$ to the population of particles $\{ \boldsymbol{\phi}_{\mathcal{M}^*}\}_g$}
\EndIf
\If {$i<N$}  {set $i=i+1$ and go to 4}
\EndIf
\If {$g < G$} 
\State{Set $g=g+1$}
\State{Update $\varepsilon_g$ \textcolor{black}{(using the median of the previous population)}}
\State{Go to 3}
\EndIf
      \label{ABCSMC}

\end{algorithmic}
\end{algorithm}%

The ABC algorithm used is the {{\it{rejection sampler}}} which is described in {\bf{Algorithm 1}}.
It should be remarked that the ABC rejection sampler is not efficient in the sense that it may lead to a high number of rejections as the generation/population advances. This pattern occurs since that the tolerance at the $g$-th population  $\varepsilon_g$ is updated  following $\varepsilon_1> \varepsilon_2 > \ldots > \varepsilon_g$. Nevertheless, it was amenable to our purposes due to the fact that the predictive models presented in Section \ref{sec:models} are analytic and run quite fast. {\textcolor{black}{More efficient sampling strategies can be found in \cite{Sisson2007} (sequential Monte Carlo, ABC-SMC) and \cite{Abdessalem2019} (nested sampling, ABC-NS)}}.

As a final note, the ABC model selection strategy penalizes over-parameterization (parsimony principle) in the Bayesian sense \cite{Beck2004,Ritto2015}. Further, if the posterior distribution of the model parameters is peaked compared to the prior distribution, then the model is also penalized. This is because a narrow peak implies that the model response is very sensitive to parameter variations. Putting in other terms, small errors in the parameter estimation will yield large errors in the model predictions.

\section{Results}\label{sec:results}


\subsection{Data}
\begin{figure}[H]
\begin{center}
\includegraphics[scale=.5]{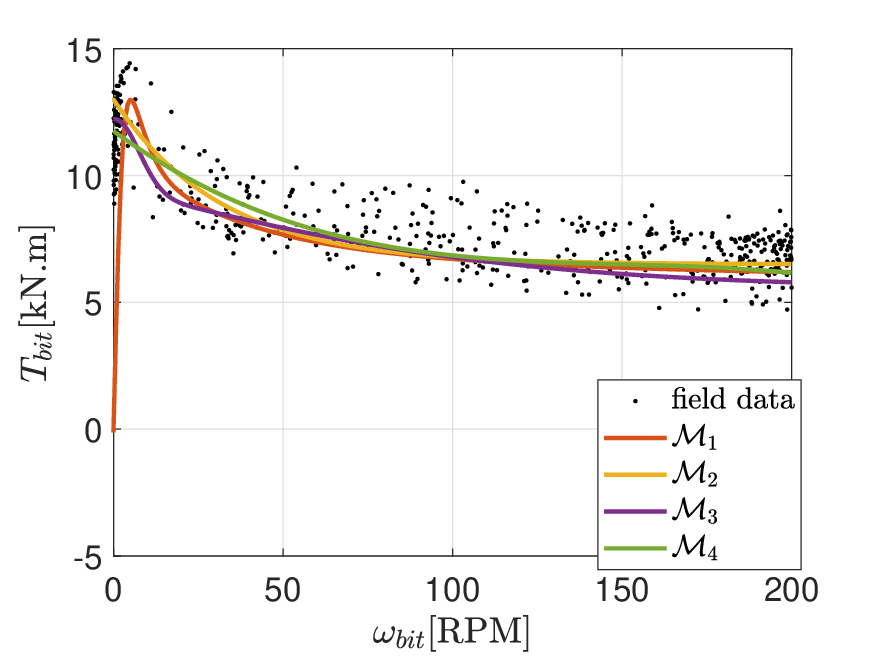}
\caption{Field data  and model predictions $\mathbf{A}(\hat{\boldsymbol{\phi}}|\mathcal{M}_k)$ computed with the least-squares estimate $\hat{\boldsymbol{\phi}}$.}
\label{fig:LSmodels}
\end{center}
\end{figure}

A set of field data from drill-string dynamics is available for model calibration and model selection, where $\mathcal{W}_{ref}=244.2$ kN. More specifically, time domain data of torque on bit $\{ T_{bit}(t_0), T_{bit}(t_1),\ldots,T_{bit}(t_{nt}) \}$ and of bit speed $\{ \omega_{bit}(t_0), \omega_{bit}(t_1),\ldots,\omega_{bit}(t_{nt}) \}$ were acquired using a downhole measurements installed at the BHA above the bit, recording at 50 Hz \cite{Shi2016}. Field data are shown in Fig.(\ref{fig:LSmodels}) by dots.

\subsection{Deterministic Models}

The parameters of the drill-string analyzed are shown in Table \ref{tb:drill_string_geometry}, where the two values of the diameters are the outer and inner dimensions. The equivalent 1-DOF system has $I_{eq}=383.33$ kgm$^2$, $\omega_n=0.85$ rad/s and $\xi=0.25$.

\begin{table}[H]
	\centering
	\caption{Geometry and parameters of the model \cite{Ritto2017}.}\label{tb:drill_string_geometry}
	\begin{tabular}{cccc}		
		\hline
		Parameter & Symbol  & Value  & Unit \\ \hline
		Shear modulus & $E$ & 85 & GPa \\
		Poisson coefficient & $\nu$ & 0.29  & -\\
		Density & $\rho$ & 7.80 $\times 10^{3}$  & $kg/m^3$ \\
		Drill-pipe length & $L_{DP}$ & 4733 & m \\
		BHA length & $L_{BHA}$ & 467& m \\	
		Drill-pipe diameter & $D_{DP}$ & 0.140 / 0.119 & m \\
		BHA diameter & $D_{BHA}$ & 0.161 /0.073 & m \\		
		\hline
	\end{tabular}
\end{table}

The first step before running the ABC was a deterministic model calibration. This was carried out considering field data shown in Fig.\ref{fig:LSmodels} and by solving the least squares problems shown next

\begin{equation}
\hat{\boldsymbol{\phi}} = \operatorname*{arg\,min}_{\boldsymbol{\phi}}  \rho(\mathbf{y}, \mathbf{A}(\boldsymbol{\phi})) 
\label{eq:LSeq}
\end{equation}
\noindent where $\hat{\boldsymbol{\phi}}$ { was determined for each model $\mathcal{M}_k$ presented in Section \ref{sec:models}. Figure \ref{fig:LSmodels} presents model predictions $\mathbf{A}(\hat{\boldsymbol{\phi}}|\mathcal{M}_k)$ computed with the least squares estimate $\hat{\boldsymbol{\phi}}$ computed as shown in Eq.(\ref{eq:LSeq}) and are presented next:
\begin{subequations}
\label{eq:Maxwell}
\begin{align}
        \mathcal{M}_1 : \hat{\boldsymbol{\phi}} = \{b_0, b_1, b_2, b_3\} = \{5.67, 0.48, 8.79, 4.56\},         \label{eq:detM1} \\
        \mathcal{M}_2 : \hat{\boldsymbol{\phi}} = \{T_{sb}, T_{cb}, G\} = \{13, 6.5, 0.3 \}, \label{eq:detM2} \\
        \mathcal{M}_3 : \hat{\boldsymbol{\phi}} = \{a_0, \ldots, a_5\} = \{ 2.72, 1, 0.09, 9.52, 4, 0.08\}, \label{eq:detM3} \\
        \mathcal{M}_4 : \hat{\boldsymbol{\phi}} = \{c_0, \ldots, c_3\} = \{11.8, -0.93, 0.057,-1.2\times10^{-3} \}, \label{eq:detM4}
\end{align}
\end{subequations}
\noindent where the model parameters shown in Eqs.(\ref{eq:detM1}-\ref{eq:detM4}) consider models with bit speed $\dot{\theta}$ in radians per second, and the torque is given in kNm.} Further, the least squares estimates $\hat{\boldsymbol{\phi}}$ shown in Eqs.(\ref{eq:detM1}-\ref{eq:detM4}) will be the basis to build the priors $\pi_0(\boldsymbol{\phi} | \mathcal{M}_k)$ required for running the ABC algorithm as shown next.

%

\subsection{Model Calibration and Model Selection}

\begin{figure}[H]
\begin{center}
\includegraphics[scale=.35]{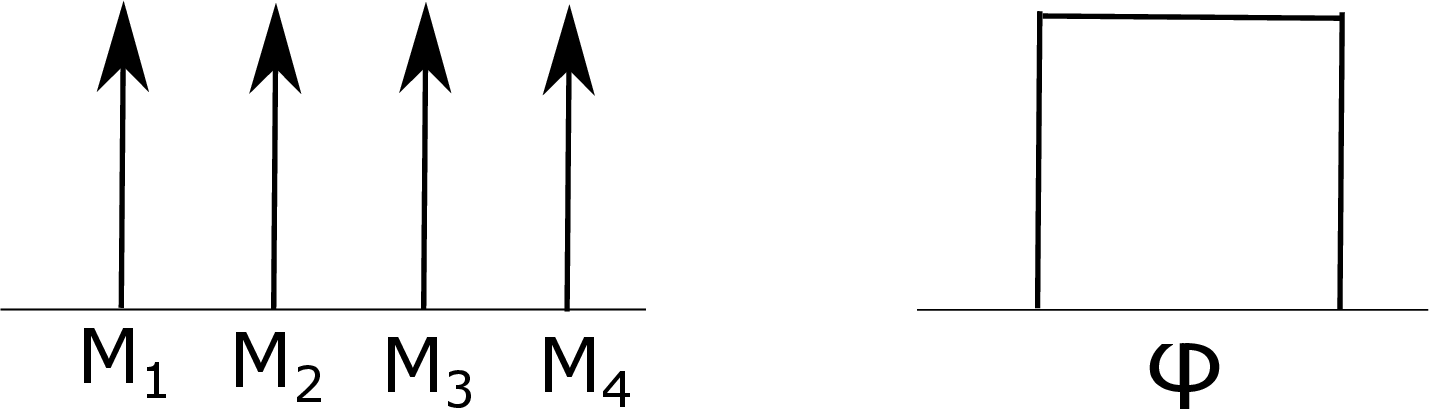}
\caption{Left: discrete Uniform prior distribution of the models. Right: continuous Uniform prior distribution of the parameters.}\label{uniform2}
\end{center}
\end{figure}

Model calibration and model selection were performed using the ABC rejection sampling shown in {\bf{Algorithm 1}}. 
No preference is assumed for any of the four models, hence the discrete Uniform distribution is chosen  for $\mathbb{P}(\mathcal{M})$, and a prior independent Uniform distribution is assumed for all  model parameters regardless of the model $\mathcal{M}_k$. See Fig. \ref{uniform2}.

Regarding the priors, the following procedure was adopted. The prior density function $\pi_0(\boldsymbol{\phi} |\mathcal{M}_k)$ assumes that $\phi_i$ and $\phi_j$ are independent random variables for all $i$ and $j$. Further,  the prior density was built as follows
\begin{equation}
\pi_0(\boldsymbol{\phi}|\mathcal{M}_k)
=\pi_0^{\delta}(\boldsymbol{\phi}|\mathcal{M}_k) = \pi_0^{\delta}({\phi}_1, \ldots, \phi_n | \mathcal{M}_k)  = \prod_{j=1}^{n} \pi_0^{\delta}(\phi_j)
\end{equation} 
\noindent where $\pi_0^{\delta}(\phi_j) \sim \mathcal{U}(\min(\phi_j), \max(\phi_j))$ is a Uniform density whose extreme values are defined as a function of the least squares estimate $\hat{\boldsymbol{\phi}}$ as shown next 

\begin{equation}
\min(\phi_j) =  \hat{\phi}_j\times (1 - \delta) \,\, \, \, \, \, \& \,\, \, \, \ \max(\phi_j) = \hat{\phi}_j\times (1 + \delta)
\label{eq:minmax}
\end{equation}
It should be  noticed that the priors were built as a function of a parameter $\delta$ which provides some flexibility to the users when assessing the sensitivity of the results provided by ABC with respect to the support of the priors as described in Eq.(\ref{eq:minmax}).  As for the model prior probability $\mathbb{P}_0(\mathcal{M}_k)$, here we adopted a less informative scenario and considered $\mathbb{P}_0(\mathcal{M}_k) = \frac{1}{4}$ for any $k \in \{1, 2, 3,4 \}$ based on the Maximum Entropy Principle. Finally, the sampling of the models along the evolution of populations considers that $\mathbb{P}(\mathcal{M}_k)=\mathbb{P}_0(\mathcal{M}_k)$ as shown in {\bf{Algorithm 1}}.

Regarding the tolerance level $\varepsilon_g$, some notes should be made. The first one is the fact that we adopted an {\it{ad-hoc}} criterion for its lower bound.  In fact, we assumed that a reasonable limit for its lower bound would be in the range [0.01, 0.02] as it would correspond to the ratio of the energy of the residue $(\mathbf{y} - \mathbf{A}(\boldsymbol{\phi}))$ to the energy of the data  $\mathbf{y}$ in the same range (see Eq.(\ref{eq:metric})). The second point refers to the rule adopted for its evolution. Here we consider that the tolerance $\varepsilon_g$ is the median of the distances computed at the previous population, i.e., $\{\rho(\mathbf{y}, \mathbf{A}(\boldsymbol{\phi}^{(i)} | \mathcal{M}^{(i)}) \}_{g-1}$. 
The simulations considered populations with $N=25\times10^3$ particles. {\textcolor{black}{The acceptance rate for $\mathcal{M}_2$ and $\mathcal{M}_3$  vary between 35 and 60\%. The other two models have an acceptance rate lower than 5\% .}}

\begin{figure}[H]
\begin{center}
\includegraphics[scale=.3]{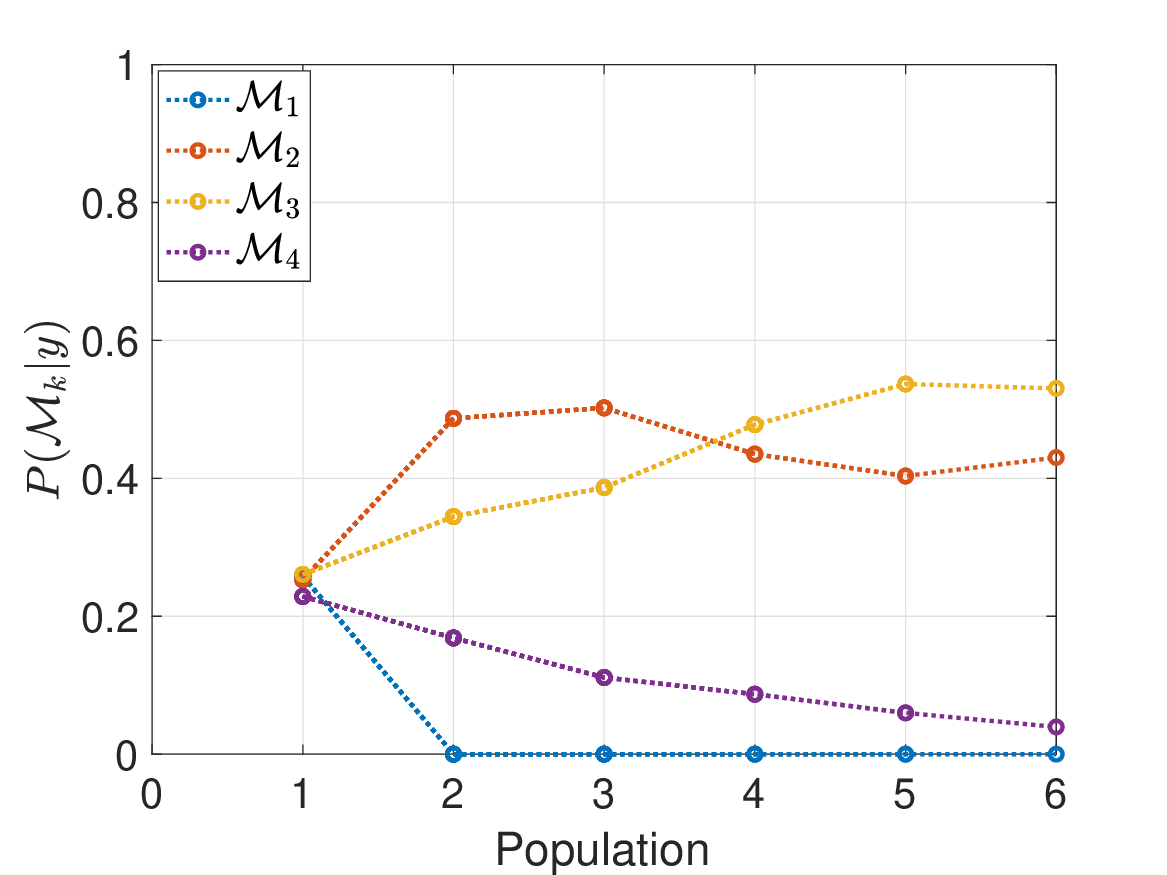}
\includegraphics[scale=.3]{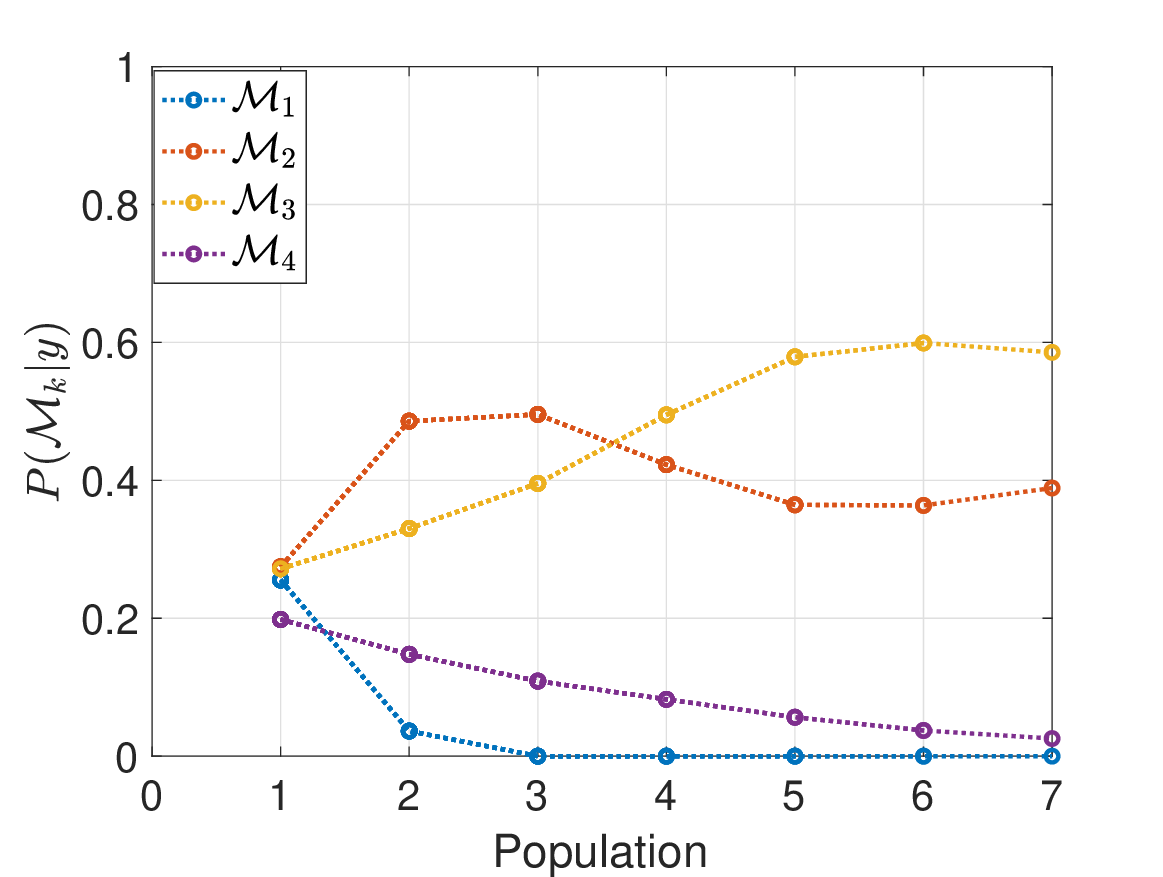}
\includegraphics[scale=.3]{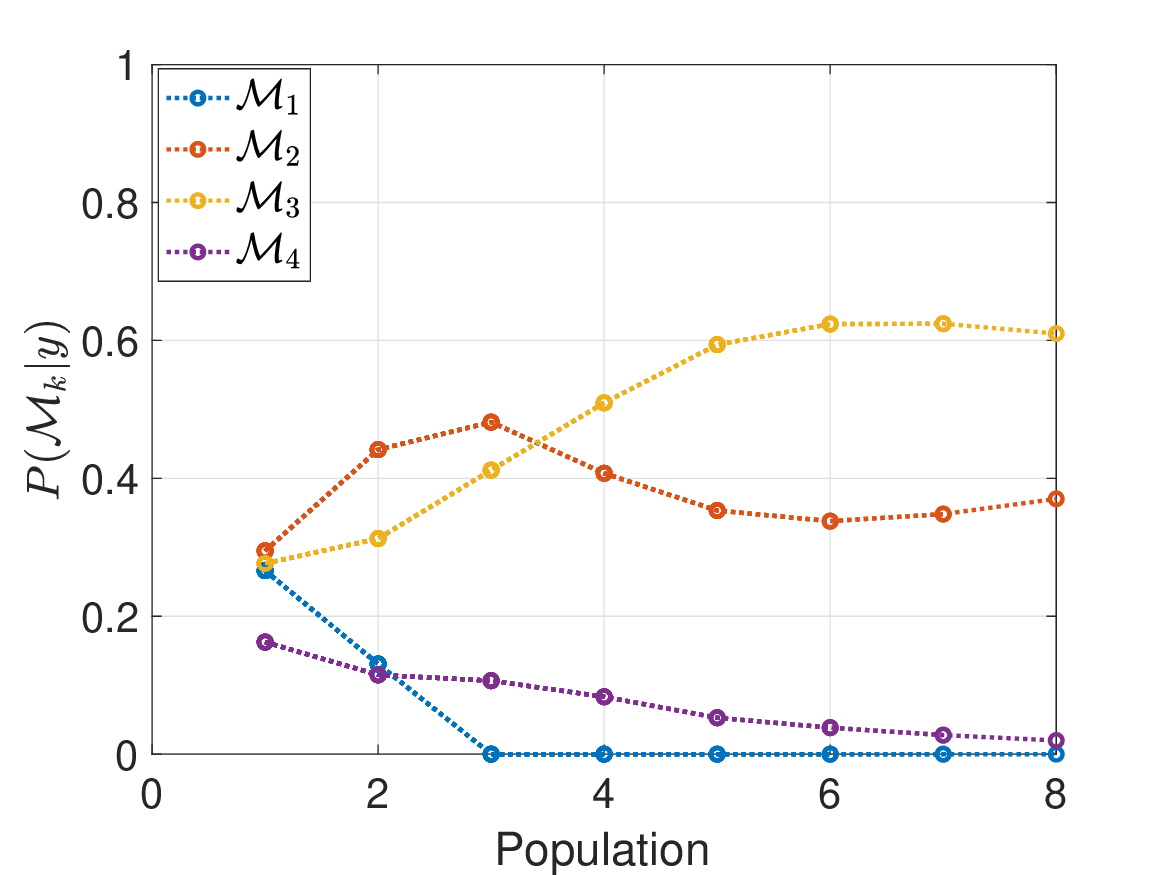}
\caption{Top left: model probability considering prior density $\pi_0^{\delta}(\boldsymbol{\phi})$ with $\delta=0.4$. Top right:  model probability considering prior density $\pi_0^{\delta}(\boldsymbol{\phi})$ with $\delta=0.6$. Bottom: model probability considering prior density $\pi_0^{\delta}(\boldsymbol{\phi})$ with $\delta=0.8$. }
\label{fig:model_prob}
\end{center}
\end{figure}

\begin{figure}[H]
\begin{center}
\includegraphics[scale=.3]{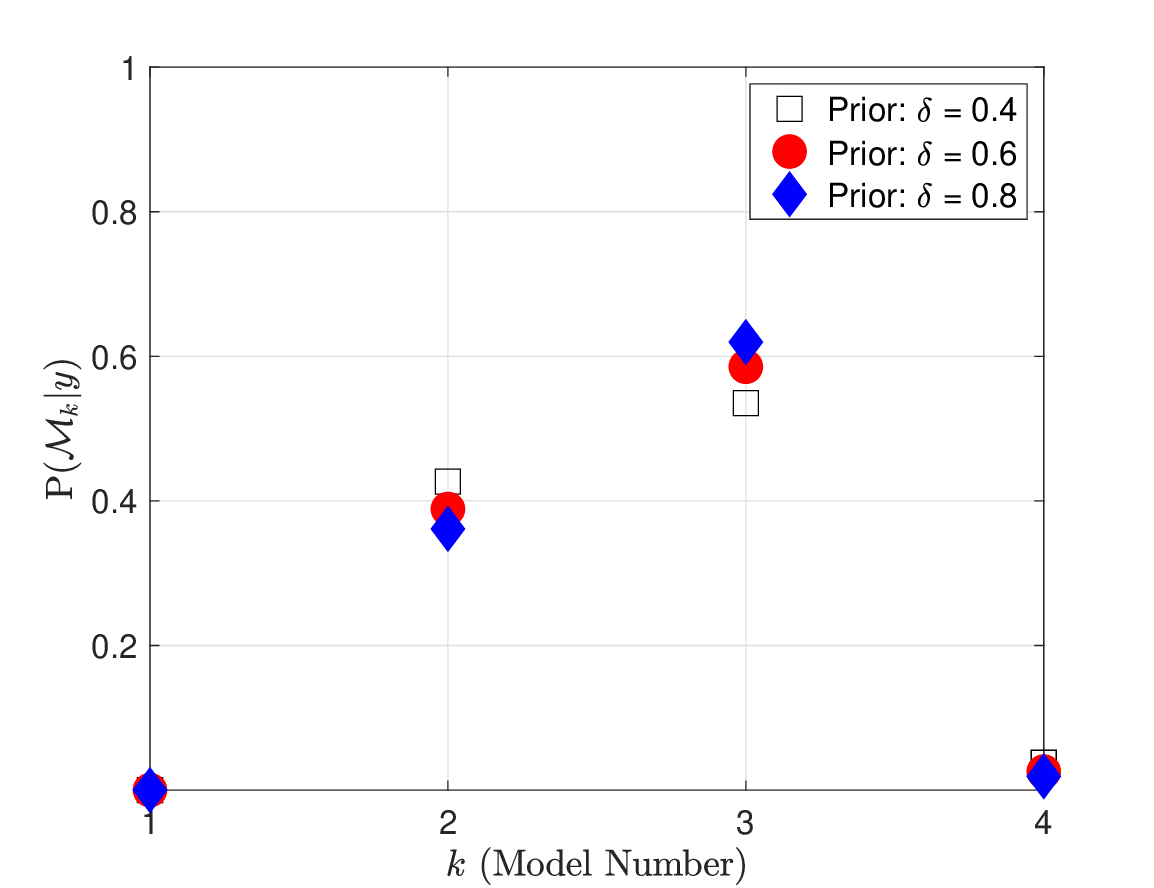}
\includegraphics[scale=.3]{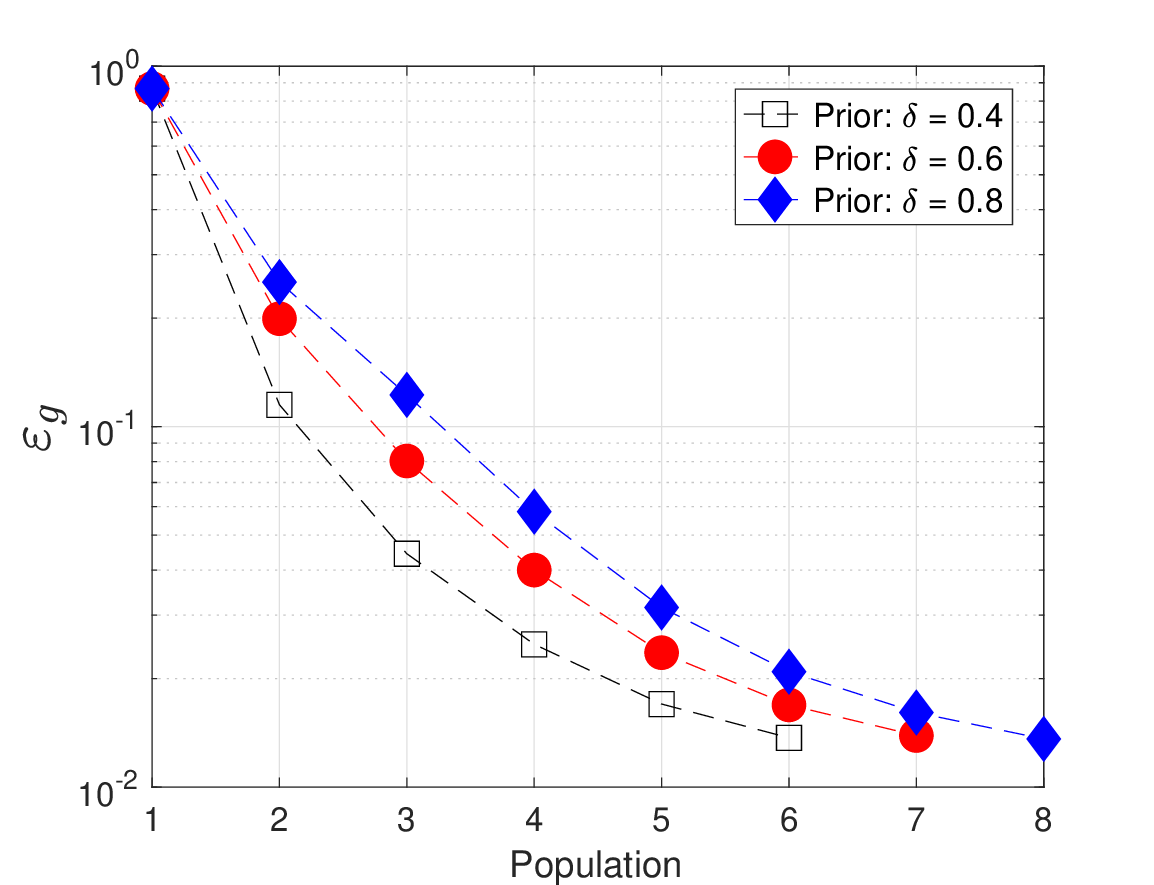}
\caption{Left: model probabilities $P(\mathcal{M}_k | \mathbf{y})$ at the last population considering different priors $\pi_0^{\delta}(\boldsymbol{\phi})$. Right: evolution of the threshold $\varepsilon_g$ along the population number.}
\label{fig:prob_last}
\end{center}
\end{figure}

Figure \ref{fig:model_prob} presents the evolution of model probabilities $\mathbb{P}(\mathcal{M}_k | \mathbf{y})$ along the populations considering different priors $\pi_0^{\delta}(\boldsymbol{\phi} | \mathcal{M}_k)$ with $\delta \in \{0.4, 0.6, 0.8 \}$.    From Fig.\ref{fig:model_prob} one observes that $\mathbb{P}(\mathcal{M}_1 | \mathbf{y})$ and $\mathbb{P}(\mathcal{M}_4 | \mathbf{y})$ are approximately zero at the last population.  Further, it is also observed that $\mathbb{P}(\mathcal{M}_3 | \mathbf{y}) > \mathbb{P}(\mathcal{M}_2 | \mathbf{y})$ for any prior under analysis. In fact, one observes the relatively low sensitivity of posterior model probabilities $\mathbb{P}(\mathcal{M}_k | \mathbf{y})$ with respect to the support of the priors described by the control parameter $\delta$ as shown in Fig.\ref{fig:prob_last}(left) which presents posterior model probabilities at the last population in a panoramic view.  Regarding the tolerance $\varepsilon$, Fig.\ref{fig:prob_last}(right)  presents its evolution  along the populations as a function of the control parameter $\delta$. From Fig.\ref{fig:prob_last}(right) one observes that although the number of populations varied with the support of the priors, the lower limit of the tolerance $\varepsilon$ remained approximately the same {\color{black}{ ($\min\{\varepsilon_g\} \approx 0.014$)}} for any prior $\pi_0^{\delta}(\boldsymbol{\phi}|\mathcal{M}_k)$.
%


%
%
%
%
%

\begin{figure}[H]
\begin{center}
\includegraphics[scale=.4]{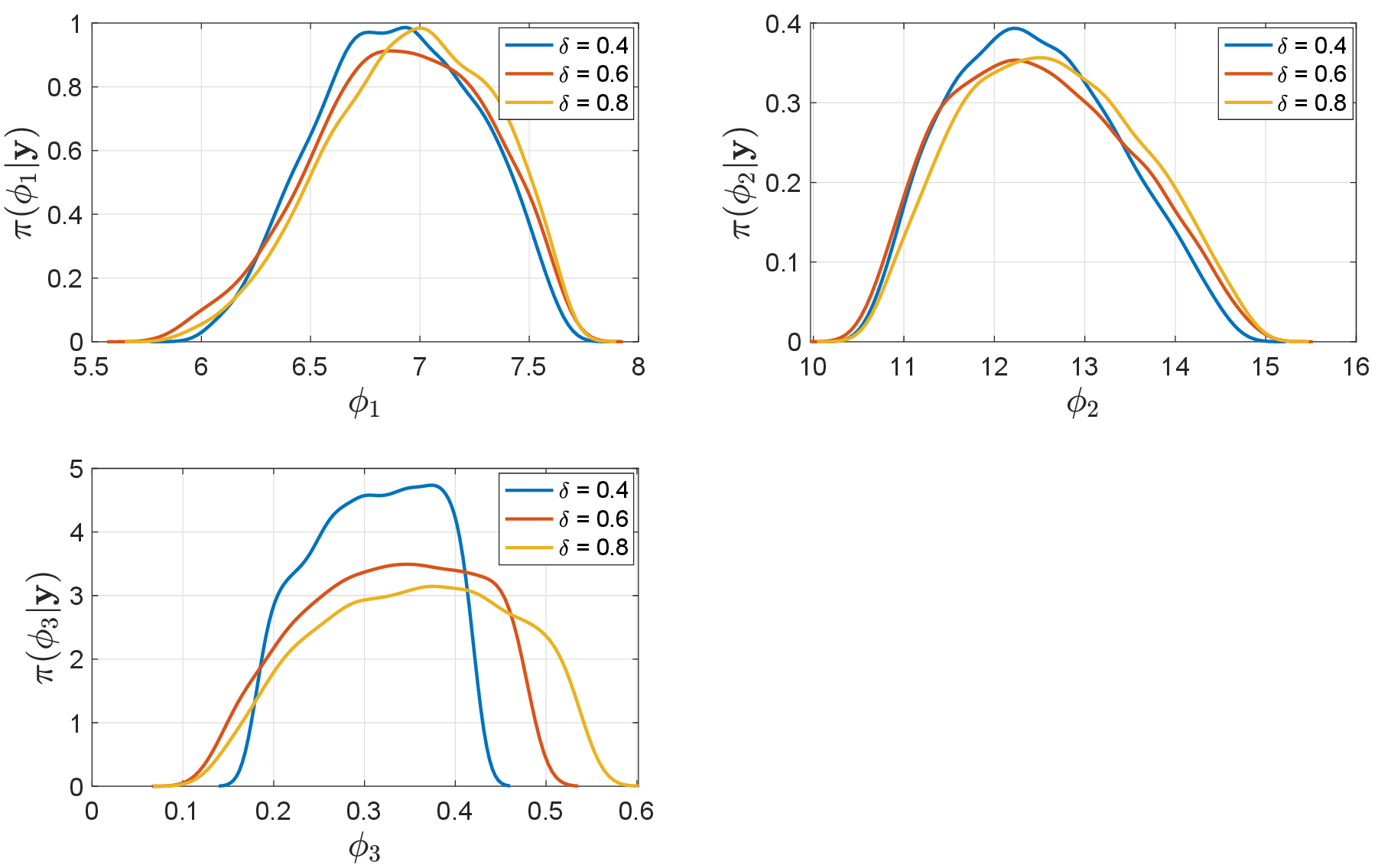}
\caption{Posterior marginal densities $\pi(\phi_i|\mathbf{y})$ for model $\mathcal{M}_2$. Note:  $\boldsymbol{\phi} = \{T_{cb}, T_{sb}, G_b  \}^{\rm T}$ for model $\mathcal{M}_2$ which is described in Eq.\ref{eq:M2}.}
\label{fig:post_M2}
\end{center}
\end{figure}

\begin{figure}[H]
\begin{center}
\includegraphics[scale=.4]{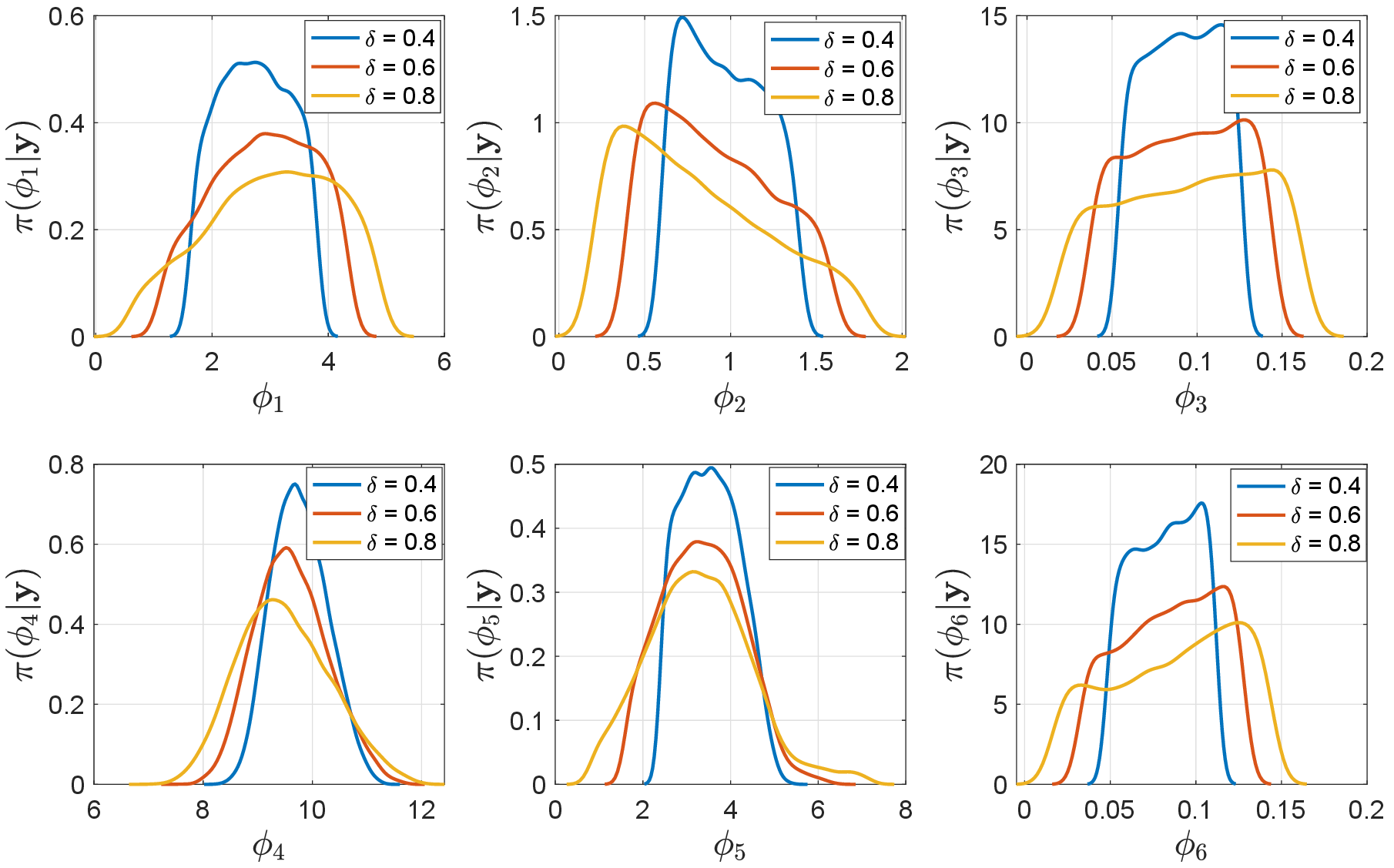}
\caption{Posterior marginal densities $\pi(\phi_i|\mathbf{y})$ for model $\mathcal{M}_3$. Note: $\boldsymbol{\phi} = \{a_0,\ldots,a_5\}^{\rm T}$  for model $\mathcal{M}_3$ which is described in Eq.\ref{eq:M3}.} 
\label{fig:post_M3}
\end{center}
\end{figure}

%
Let us analyze in details the posterior marginals $\pi(\phi_j | \mathbf{y}, \mathcal{M}_2)$ and $\pi(\phi_j | \mathbf{y}, \mathcal{M}_3)$ as $\mathcal{M}_2$ and $\mathcal{M}_3$ were the models selected by ABC as shown in Figs.\ref{fig:model_prob} and \ref{fig:prob_last}. Figure \ref{fig:post_M2} presents the posterior marginals for model $\mathcal{M}_2$ and Fig.\ref{fig:post_M3} the posterior marginals for model $\mathcal{M}_3$. For the two first parameters of $\mathcal{M}_2$ there is no noticeable difference between the posteriors when varying the priors support. {\color{black}{On the other hand}}, for the third parameter of $\mathcal{M}_2$ and all six parameters of $\mathcal{M}_3$ one can see a pattern. The posterior is more informative, i.e., thinner densities, for more informative priors, but the differences are not so significant. The density shapes are similar and the mean values change a little.


%


\begin{figure}[H]
\begin{center}
\includegraphics[scale=.7]{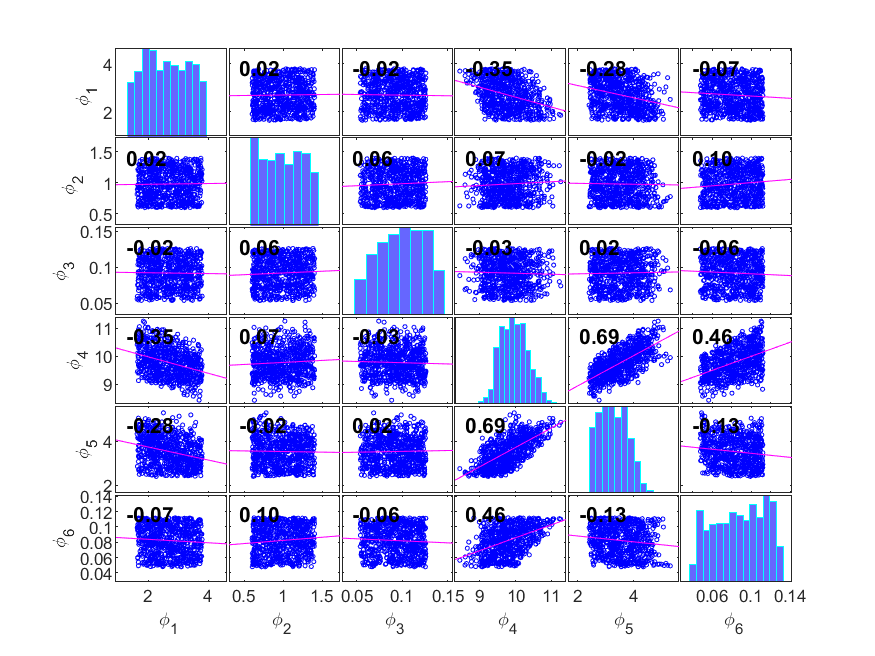}
\caption{Correlation plot for the parameters of model $\mathcal{M}_3$ with $\delta=0.4$.} 
\label{fig:m1_correlations}
\end{center}
\end{figure}

{\textcolor{black}{It is also possible to compute correlations between random variables. Figure \ref{fig:m1_correlations} shows a correlation plot for each pair of parameters  of $\mathcal{M}_3$ (Pearson's linear correlation coefficient) with the normalized histograms (700 particles). The fourth and fifth parameters (related to the $\tanh$ term) are the most correlated ones.}}

Regarding model predictions, Fig.\ref{fig:predictions} presents predictions $\mathbf{A}(\mathbf{\phi}^{(j)} | \mathcal{M}_k)$ provided by models $\mathcal{M}_2$ and $\mathcal{M}_3$ computed with posterior samples  $\boldsymbol{\phi}^{(j)} \sim \pi(\boldsymbol{\phi} | \mathbf{y})$. Black dots are field data used for validation and red circles correspond to field data used for the calibration procedure. {\textcolor{black}{The 98\% probabilistic envelopes (filled region) of both models are able to encompass a significant part of the validation data (black dots), showing that the models are consistent to represent the analyzed phenomenon. As for the tolerance $\varepsilon_g$, for $\varepsilon_7=0.014$ (7th population), the envelopes are too thin compared to data fluctuation as can be seen in Fig.\ref{fig:predictions}. A better choice would be  $\varepsilon_4=0.040$  (4th population), which presents an excellent fit.
The main difference between the two stochastic models is observed close to the static torque. The statistical envelope of $\mathcal{M}_3$ has a better fit (ranging from about 10 to 16 kNm), while it is wider for $\mathcal{M}_2$ (ranging from about 8 to 17 kNm). Points  in the range of $[50,120]$ RPM escape a little from the predictions provided by the models. One possible explanation for the misfit is that the weight on bit is not constant in the real operation, and the models do not consider axial force fluctuations. }

\begin{figure}[H]
\begin{center}
\includegraphics[scale=.4]{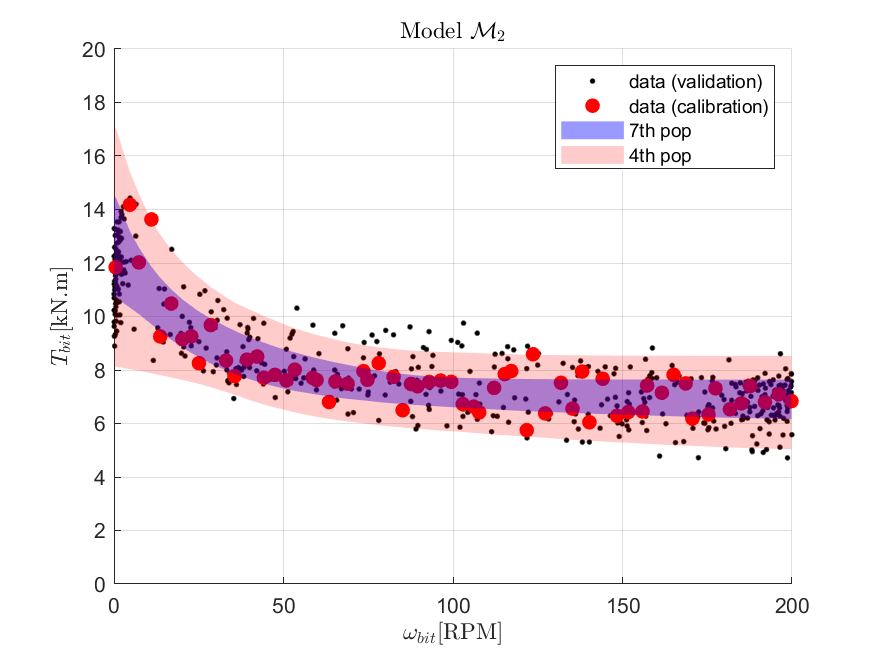}
\includegraphics[scale=.4]{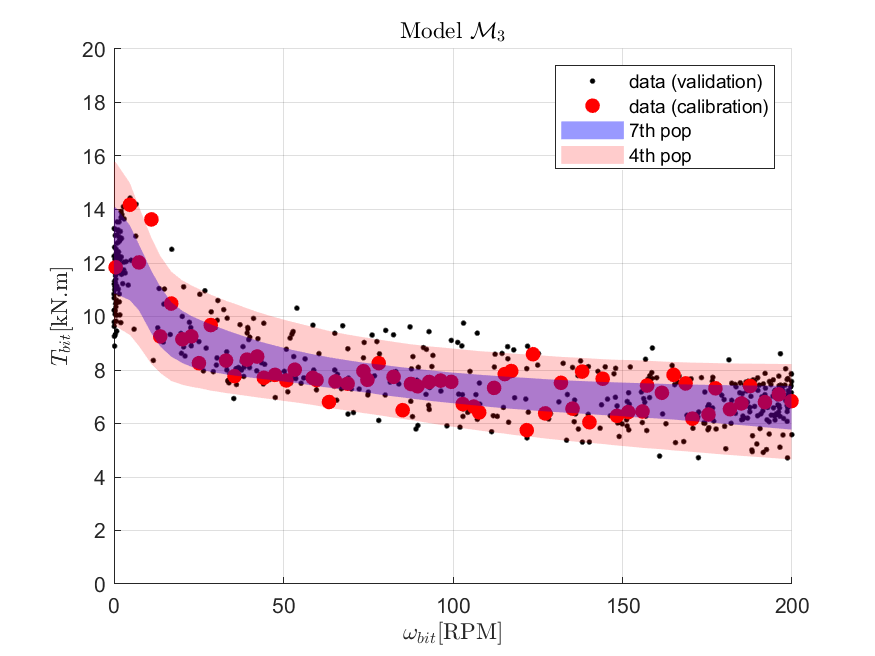}
\caption{Left: Model predictions $\mathbf{A}(\mathbf{\phi} | \mathcal{M}_2)$ provided by $\pi(\boldsymbol{\phi} | \mathbf{y}, \mathcal{M}_2)$ in gray lines. Right: Model predictions $\mathbf{A}(\mathbf{\phi} | \mathcal{M}_3)$ provided by $\pi(\boldsymbol{\phi} | \mathbf{y}, \mathcal{M}_3)$ in gray lines. The gray region represents the 98\% probabilistic envelope, blue dots are field data used for validation, and red circles correspond to field data used for the calibration procedure. \textcolor{black}{Note: $\varepsilon_7 = 0.014$}}
\label{fig:predictions}
\end{center}
\end{figure}





\textcolor{black}{The drill-bit faces different soil characteristics along the way. Whenever new data are available, the procedure should be repeated: least-squares estimate and ABC model selection computed. Then, a stochastic stability map (see next section) can be constructed with the most suited model (or a mix of the best models) to support engineering and operational decisions. 
}

\subsection{Stability analysis}

In this section, a stability map is built varying the values of the operational parameters {\color{black}{angular speed of the rotary table and weight on bit, namely:}} $(\Omega,\mathcal{W})$. The procedure is the following. The system is linearized as explained in Section \ref{sec:stability}. If the real part of all eigenvalues of the system matrix $\textbf{A}$ is negative the system is stable, otherwise it is unstable. This result is used to construct stability boundary curves.

Figure \ref{boundary_modelsID2} shows the stability map using the identified (MAP) parameters. The colored curves represent the bifurcation boundary for each model: it separates the stable region (points to the right of the curve) from the unstable region (points to the left of the curve). This is a well-known stability behavior for the torsional dynamics of a drill-string \cite{Nogueira2018}.

\ref{sec:appendix} shows that the stability map is equivalent if a FE torsional model is considered. The system tends to instability as the weight on bit increases and the angular speed at the top decreases. To be more specific, it represents a Hopf bifurcation \cite{Vlajic2017102,Depouhon2014,Vromen2017,Yan2019,Zheng2020,CunhaLima2015}, where the fixed point loses stability, as a pair of complex conjugate eigenvalues crosses the complex plane imaginary axis; hence a limit cycle branches from the fixed point. 

It is interesting to note that $\mathcal{M}_1$ presents the most optimistic behavior (larger stability region) and $\mathcal{M}_3$ becomes more conservative as $\Omega$ increases. The shape of these boundaries depends solely on the bit-rock interaction model, i.e., on how the torque-on-bit varies with the bit speed.

\begin{figure}[H]
\begin{center}
\includegraphics[scale=.5]{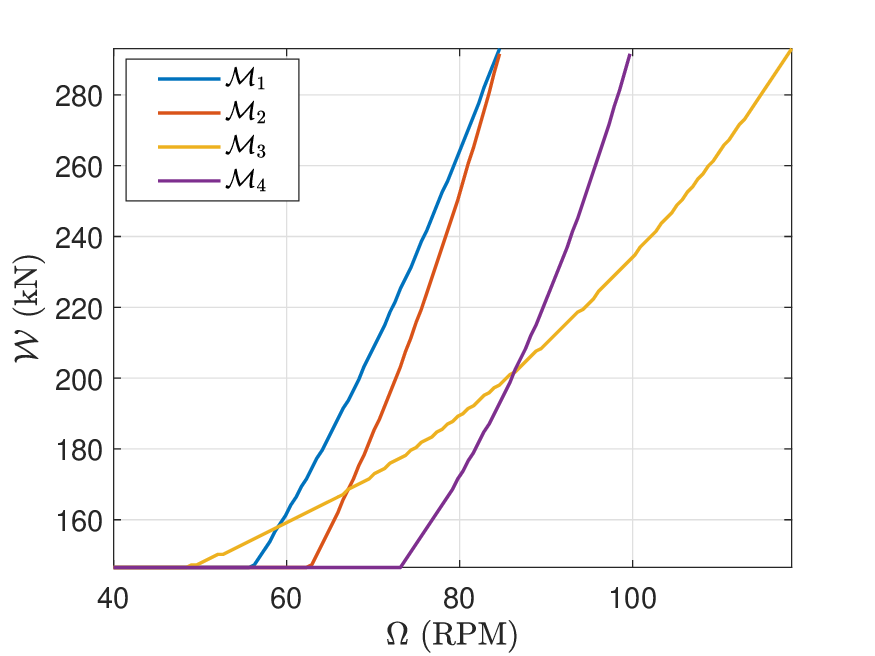}
\caption{Stability map obtained with the MAP values for the model parameters. Operational points ($\Omega,\mathcal{W}$) in the region above the curves are unstable and below are stable.}\label{boundary_modelsID2}
\end{center}
\end{figure}

We can also build a stochastic stability map using samples from the posterior distribution (700 samples from case 1); see Fig. \ref{Estbilidade_estocastica}. The solid lines in Fig. \ref{Estbilidade_estocastica} represent the deterministic case using the MAP parameter values. $\mathcal{M}_1$ and $\mathcal{M}_4$ are not in the figure because they did not perform as well as the others. The models with higher probability are $\mathcal{M}_2$ (40\%) and $\mathcal{M}_3$ (60\%), and they yield very different curves in the stability map. $\mathcal{M}_3$ is clearly more conservative than $\mathcal{M}_2$, generating a larger instability region (area to the left of the curve). 

The dashed lines in Fig.  \ref{Estbilidade_estocastica} gather the stochastic results, representing the limit for which the probability of instability is lower than 2\%. The stochastic results are to the right of the deterministic ones since they take into account uncertainties and can be thought of as a safety factor in the analysis. Finally, the black dashed line is calculated mixing $\mathcal{M}_2$ with $\mathcal{M}_3$ (stochastic models), weighing 40\% to $\mathcal{M}_2$ and 60\% to $\mathcal{M}_3$. This curve is a compromise between $\mathcal{M}_2$ and $\mathcal{M}_3$: more conservative than $\mathcal{M}_2$ and less than $\mathcal{M}_3$. Note that the black curve is drawn until the maximum speed is observed for $\mathcal{M}_2$; after that point, there is no more information about this model. 


\begin{figure}[H]
\begin{center}
\includegraphics[scale=.5]{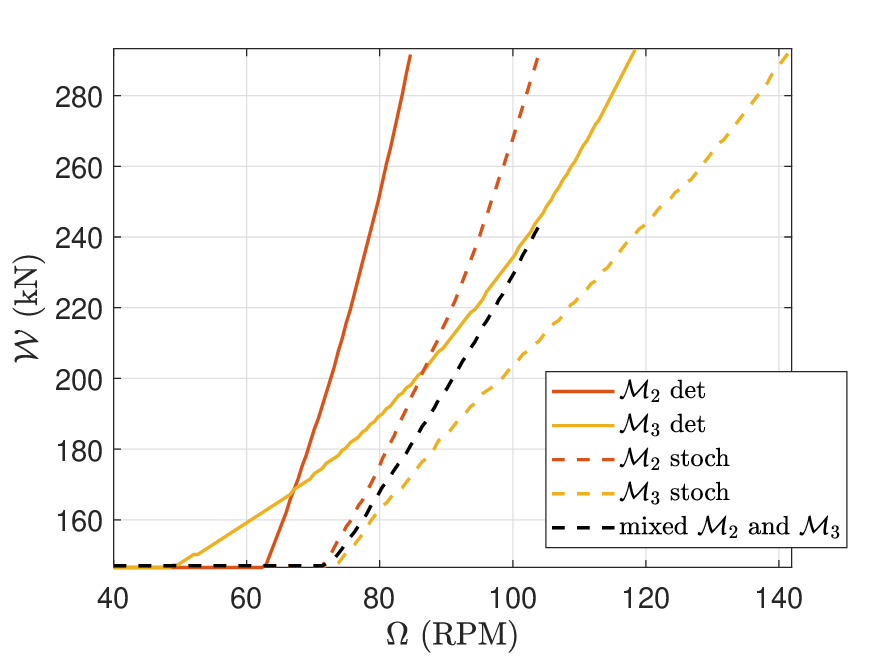}
\caption{Linear stability map obtained with the MAP values for the model parameters (continuous line) and for a percentile of 2\% of the stochastic system (dashed lines). The black dashed line is a mix between the stochastic results of $\mathcal{M}_2$ and $\mathcal{M}_3$. Operational points ($\Omega,\mathcal{W}$) in the region above the curves are unstable and below are stable.}\label{Estbilidade_estocastica}
\end{center}
\end{figure}

\section{Concluding remarks}\label{sec:conclusions}

An approximate Bayesian computation (ABC) strategy was used for model selection and parameter calibration of bi-rock interaction models. The new proposed bit-rock interaction model that combines the hyperbolic tangent function with the exponential function performed better than the other three models, considering the data available. It should be noted that all the models have a good correspondence with the data, and the results might change if different data are analyzed or different prior knowledge is given.

A torsional stability map was built for the deterministic and stochastic systems. It was observed that the stability boundary of the different models varies significantly. Furthermore, the constructed stochastic boundary pushes the bound to the right, decreasing the stable region and serving as a safety factor.
\ref{sec:appendix} shows a finite element (FE) formulation for the torsional drill-string and concludes that the torsional stability map is equivalent to the one obtained with a single degree of freedom system if the damping parameters are adjusted accordingly.


The proposed ABC strategy should be used for more elaborated bit-rock interaction models (e.g. coupled axial-torsional \cite{Tucker2003,Richard2007}) and different data. In the future we aim to use a recently developed strategy for model selection and parameter calibration, that combines reinforcement learning with ABC \cite{Ritto2022}, to speed up the model selection in real-time applications. The idea is to apply reinforcement learning to reinforce the best models choosing them more frequently than the least adapted ones.
	
	\section*{Acknowledgment}
	
	We  would like to acknowledge that this investigation was financed in part by the Brazilian agencies: \text{Coordenação de Aperfeiçoamento de Pessoal de Nível Superior} (CAPES) - Finance code 001 - Grant PROEX 803/2018, \textit{Conselho Nacional de Desenvolvimento Cient\'ifico e Tecnol\'ogico} (CNPQ) - Grant number 312355/2020-3 - and \textit{Funda\c{c}\~ao Carlos Chagas Filho de Amparo \`a Pesquisa do Estado do Rio de Janeiro} (FAPERJ) - Grant E-26/201.183/2022.	

	\bibliography{bibdtwin.bib}

\begin{thebibliography}{10}
\expandafter\ifx\csname url\endcsname\relax
  \def\url#1{\texttt{#1}}\fi
\expandafter\ifx\csname urlprefix\endcsname\relax\def\urlprefix{URL }\fi
\expandafter\ifx\csname href\endcsname\relax
  \def\href#1#2{#2} \def\path#1{#1}\fi

\bibitem{Kreuzer2012}
E.~Kreuzer, M.~Steidl, Controlling torsional vibrations of drill strings via
  decomposition of traveling waves, Archive of Applied Mechanics 82~(4) (2012)
  515--531.

\bibitem{Vromen2017}
T.~Vromen, N.~van de Wouw, A.~Doris, P.~Astrid, H.~Nijmeijer, Nonlinear
  output-feedback control of torsional vibrations in drilling systems,
  International Journal of Robust and Nonlinear Control 27~(17) (2017)
  3659--3684.

\bibitem{Monteiro2017}
H.~Monteiro, M.~Trindade, Performance analysis of proportional-integral
  feedback control for the reduction of stick-slip-induced torsional vibrations
  in oil well drillstrings, Journal of Sound and Vibration 398 (2017) 28--38.

\bibitem{Ritto2017}
T.~G. Ritto, R.~R. Aguiar, S.~Hbaieb, Validation of a drill string dynamical
  model and torsional stability, Meccanica 52~(11-12) (2017) 2959--2967.

\bibitem{Tian2019}
J.~Tian, G.~Li, L.~Dai, L.~Yang, H.~He, S.~Hu, Torsional vibrations and
  nonlinear dynamic characteristics of drill strings and stick-slip reduction
  mechanism, Journal of Computational and Nonlinear Dynamics 14~(8) (2019).

\bibitem{Abbassian1998}
F.~Abbassian, V.~Dunayevsky, Application of stability approach to torsional and
  lateral bit dynamics, SPE Drilling and Completion 13~(2) (1998) 99--107.

\bibitem{Yigit1998}
A.~Yigit, A.~Christoforou, Coupled torsional and bending vibrations of
  drillstrings subject to impact with friction, Journal of Sound and Vibration
  215~(1) (1998) 167--181.

\bibitem{Volpi2021}
L.~P. Volpi, D.~M. Lobo, T.~G. Ritto, A stochastic analysis of the coupled
  lateral–torsional drill string vibration, Nonlinear Dynamics 103~(1) (2021)
  49--61.

\bibitem{Richard2007}
T.~Richard, C.~Germay, E.~Detournay, A simplified model to explore the root
  cause of stick-slip vibrations in drilling systems with drag bits, Journal of
  Sound and Vibration 305~(3) (2007) 432--456.

\bibitem{Lobo2020}
D.~M. Lobo, T.~G. Ritto, D.~A. Castello, M.~L.~M. Souza, On the calibration of
  drill-string models based on hysteresis cycles data, International Journal of
  Mechanical Sciences 117 (2020) 105578.

\bibitem{Tucker1999}
R.~Tucker, C.~Wang, An integrated model for drill-string dynamics, Journal of
  Sound and Vibration 224~(1) (1999) 123--165.

\bibitem{Khulief2005}
Y.~Khulief, H.~Al-Naser, Finite element dynamic analysis of drillstrings,
  Finite Elements in Analysis and Design 41~(13) (2005) 1270--1288.

\bibitem{Ritto2009}
T.~Ritto, C.~Soize, R.~Sampaio, Non-linear dynamics of a drill-string with
  uncertain model of the bit-rock interaction, International Journal of
  Non-Linear Mechanics 48~(8) (2009) 865--876.

\bibitem{deMoraes2019}
L.~de~Moraes, M.~Savi, Drill-string vibration analysis considering an
  axial-torsional-lateral nonsmooth model, Journal of Sound and Vibration 438
  (2019) 220--237.

\bibitem{Ritto2018}
T.~G. Ritto, M.~Ghandchi-Tehrani, Active control of stick-slip torsional
  vibrations in drill-strings, Journal of Vibration and Control 25~(1) (2018)
  194--202.

\bibitem{Detournay1992}
E.~Detournay, P.~Defourny, A phenomenological model for the drilling action of
  drag bits, International Journal of Rock Mechanics and Mining Sciences and
  29~(1) (1992) 13--23.

\bibitem{Tucker2003}
R.~Tucker, C.~Wang, Torsional vibration control and cosserat dynamics of a
  drill-rig assembly, Meccanica 38~(1) (2003) 143--159.

\bibitem{Tucker97}
R.~W. Tucker, C.~Wang, {The excitation and control of torsional slip-stick in
  the presence of axial vibrations.},
  http://www.lancs.ac.uk/users/SPC/Physics.htm (1997).

\bibitem{Navarro2004}
E.~Navarro-Lopez, R.~Suarez, Practical approach to modelling and controlling
  stick-slip oscillations in oilwell drillstrings, in: Proceedings of the 2004
  IEEE International Conference on Control Applications, Vol. 8281780, 2004,
  pp. 1454--1460.

\bibitem{Real2018}
F.~Real, A.~Batou, T.~Ritto, C.~Desceliers, R.~Aguiar, Hysteretic bit/rock
  interaction model to analyze the torsional dynamics of a drill string,
  Mechanical Systems and Signal Processing 111 (2018) 222--233.

\bibitem{Kapitaniak2016}
M.~Kapitaniak, V.~Hamaneh, M.~Wiercigroch, Torsional vibrations of helically
  buckled drill-strings: Experiments and fe modelling, Journal of Physics:
  Conference Series 721~(1) (2016) 12012.

\bibitem{Aldrich1997}
J.~Aldrich, R. a. fisher and the making of maximum likelihood 1912 - 1922,
  Statistical Science 12~(3) (1997) 162--176.

\bibitem{Ritto2010}
T.~G. Ritto, C.~Soize, R.~Sampaio, Probabilistic model identification of the
  bit–rock-interaction-model uncertainties in nonlinear dynamics of a
  drill-string, Mechanics Research Communications 37~(6) (2010) 584--589.

\bibitem{Kaipio2006}
J.~Kaipio, E.~Somersalo, Statistical and Computational Inverse Problems,
  Springer, 2004.

\bibitem{Ritto2015b}
T.~G. Ritto, {B}ayesian approach to identify the bit–rock interaction
  parameters of a drill-string dynamical model, Journal of the Brazilian
  Society of Mechanical Sciences and Engineering 37~(4) (2015) 1173--1182.

\bibitem{Nogueira2018}
B.~F. Nogueira, T.~G. Ritto, {Stochastic torsional stability of an oil
  drill-string}, Meccanica 53~(11-12) (2018) 3047--3060.

\bibitem{Chiachio2014}
M.~Chiachio, J.~Beck, J.~Chiachio, G.~Rus, Approximate {B}ayesian computation
  by subset simulation, SIAM J. Sci. Comput. 36~(3) (2014) A1339--A1358.

\bibitem{Abdessalem2018}
A.~Ben~Abdessalem, N.~Dervilis, D.~Wagg, K.~Worden, Model selection and
  parameter estimation in structural dynamics using approximate {B}ayesian
  computation, Mechanical Systems and Signal Processing 99 (2018) 306--325.

\bibitem{Ritto2022}
T.~G. Ritto, S.~Beregi, D.~A.~W. Barton, Reinforcement learning and approximate
  {B}ayesian computation for model selection and parameter calibration applied
  to a nonlinear dynamical system, Mechanical Systems and Signal Processing
  accepted for publication (2022).

\bibitem{Toni2008}
T.~Toni, D.~Welch, N.~Strelkowa, A.~Ipsen, M.~P. Stumpf, Approximate {B}ayesian
  computation scheme for parameter inference and model selection in dynamical
  systems 6 (2009) 187--202.

\bibitem{Abdessalem2019}
A.~B. Abdessalem, N.~Dervilis, D.~Wagg, K.~Worden, Model selection and
  parameter estimation of dynamical systems using a novel variant of
  approximate {B}ayesian computation, Mechanical Systems and Signal Processing
  122 (2019) 364--386.

\bibitem{ABCHandbook}
S.~A. Sisson, Y.~Fan, M.~A. Beaumont, Overview of ABC, Chapter 1 in Handbook of
  Approximate {B}ayesian Computation Edited by S.A. Sisson, Y. Fan and M.A.
  Beaumont, Chapman \& Hall/CRC, 2019.

\bibitem{Beck2004}
J.~Beck, K.-V. Yuen, Model selection using response measurements: {B}ayesian
  probabilistic approach, Journal of Engineering Mechanics 130~(2) (2004)
  192--203.

\bibitem{Ritto2015}
T.~Ritto, L.~Nunes, {B}ayesian model selection of hyperelastic models for
  simple and pure shear at large deformations, Computers and Structures 156
  (2015) 101--109.

\bibitem{Castello2018}
W.~P. Hernandéz, D.~A. Castello, C.~F.~T. Matt, On the model building for
  transmission line cables: a bayesian approach, Inverse Problems in Science
  and Engineering 26(12) (2018) 1784--1812.

\bibitem{Sisson2007}
S.~Sisson, Y.~Fan, M.~Tanaka, Sequential monte carlo without likelihoods,
  Proceedings of the National Academy of Sciences of the United States of
  America 104~(6) (2007) 1760--1765.

\bibitem{Shi2016}
J.~Shi, B.~Durairajan, R.~Harmer, W.~Chen, F.~Verano, Y.~Arevalo, C.~Douglas,
  T.~Turner, D.~Trahan, J.~Touchet, Y.~Shen, A.~Zaheer, F.~Pereda,
  K.~Robichaux, D.~Cisneros, Integrated efforts to understand and solve
  challenges in 26-in salt drilling, gulf of mexico, 2016.

\bibitem{Vlajic2017102}
N.~Vlajic, A.~Champneys, B.~Balachandran, Nonlinear dynamics of a jeffcott
  rotor with torsional deformations and rotor-stator contact, International
  Journal of Non-Linear Mechanics 92 (2017) 102--110.

\bibitem{Depouhon2014}
A.~Depouhon, E.~Detournay, Instability regimes and self-excited vibrations in
  deep drilling systems, Journal of Sound and Vibration 333~(7) (2014)
  2019--2039.

\bibitem{Yan2019}
Y.~Yan, M.~Wiercigroch, Dynamics of rotary drilling with non-uniformly
  distributed blades, International Journal of Mechanical Sciences 160 (2019)
  270--281.

\bibitem{Zheng2020}
X.~Zheng, V.~Agarwal, X.~Liu, B.~Balachandran, Nonlinear instabilities and
  control of drill-string stick-slip vibrations with consideration of
  state-dependent delay, Journal of Sound and Vibration 473 (2020).

\bibitem{CunhaLima2015}
L.~C. Cunha-Lima, R.~R. Aguiar, T.~G. Ritto, S.~Hbaieb, Analysis of the
  torsional stability of a simplified drillstring, in: Proceedings of the 17th
  International Symposium on Dynamic Problems of Mechanics, 2015.

\bibitem{Ritto2022b}
T.~Ritto, K.~Worden, D.~Wagg, F.~Rochinha, P.~Gardner, A transfer
  learning-based digital twin for detecting localised torsional friction in
  deviated wells, Mechanical Systems and Signal Processing 173 (2022).

\end{thebibliography}
	
\appendix
\section{Finite element model and stability analysis}\label{sec:appendix}

This appendix shows a finite element (FE) formulation for the torsional drill-string \cite{Ritto2015b,Ritto2022b} and concludes that the stability map is equivalent to the one obtained with a single degree of freedom system. The mass and stiffness elementary FE matrices, with $l_{\text{el}}$ the element length, are given by:

\begin{equation}
\textbf{M}_{\text{el}} =\rho J l_{\text{el}}
\left[\begin{array}{cc}
      1/3   & 1/6 \\
      1/6   & 1/3 
    \end{array}\right]\quad,\quad \textbf{K}_{\text{el}} =\frac{G J}{l_{\text{el}}}
\left[\begin{array}{cc}
     1   & -1 \\
      -1   & 1 
    \end{array}\right]\label{elem_matrices}\,,
\end{equation}

\noindent where $\rho$ is the material density, $G$ the shear modulus, and the area moment of inertia $J$ is different for the drill-pipe DP and the BHA,

\begin{equation}
J_{\text{DP}} = \frac{\pi}{32} (D_{\text{DPo}}^4 - D_{\text{DPi}}^4)\quad,\quad
J_{\text{BHA}} = \frac{\pi}{32}(D_{\text{BHAo}}^4 - D_{\text{BHAi}}^4)\,,
\end{equation}

\noindent where the outer $o$ and inner $i$ radius are considered in the above expressions. The finite element discretized system, without the source terms, is written as:

\begin{equation}
\label{eq:EF_EQMOT}
    \textbf{M}\ddot{\textbf{u}}(t)+\textbf{C}\dot{\textbf{u}}(t)+\textbf{K}{\textbf{u}}(t)=\textbf{f}(\dot{\textbf{u}}(t))\,,
\end{equation}

\noindent where $\textbf{f}(\dot{\textbf{u}}(t))=(0,...,0,-T_{bit}(\dot{\theta}_{n_{el}}))$ is the vector that contains the bit-rock interaction torque (a boundary condition), $\textbf{M}$, $\textbf{C}$ and $\textbf{K} \in \mathbb{R}^{n_{\text{el}} \times n_{\text{el}}}$ are the mass, damping and stiffness matrices of the system, discretized in $n_{\text{el}}$ finite elements \cite{Ritto2015b}. The response vector $\textbf{u}(t)=(\theta_1,...,\theta_{n_{el}})$, and $\textbf{x}(t)=(\textbf{u}(t),\dot{\textbf{u}}(t))$.

The boundary condition $\textbf{u}(0)=0$ is used because we assume a constant angular speed at the top; thus the first angular degree of freedom is not free to oscillate. \textcolor{black}{A common strategy to impose the constant $\Omega$ at the top of the system is to move the corresponding terms to the right side of the equation \cite{Ritto2009}; this is not needed here since we only care about stability}. Finally, the damping matrix takes into account different mechanisms, such as structural and viscous damping due to the drilling mud, and a proportional model is employed $\textbf{C}=\alpha \textbf{M} + \beta \textbf{K}$ ($\alpha,\beta \in \mathbb{R}^+$).}

To compute the stability of the system, the eigenvalues of the Jacobean matrix $\textbf{A}\in \mathbb{R}^{2n_{\text{el}}\times 2n_{\text{el}}}$ is computed. The system is linearized around the rotating frame, i.e., \textcolor{black}{ $\textbf{x}^*=(\Omega t-\theta_{01},\Omega t-\theta_{02},\ldots,\Omega t-\theta_{0n_{el}},\Omega,...,\Omega)\in \mathbb{R}^{2n_{el}}$. The angles $\theta_{01}$, $\theta_{02}$, etc. are necessary for dynamic equilibrium; in this configuration, the constant torque at the bit balances the elastic forces, and the column rotates with no torsional oscillations.}

We want to evaluate the system stability for different values of the pair $(\Omega,\mathcal{W})$; the constant angular speed at the top $\Omega$ and weight on bit $\mathcal{W}$. {\color{black}{Based on Eq.(\ref{eq:EF_EQMOT}), the state matrix \textbf{A} (Jacobian) is given as shown next}}




\begin{equation}
    \textbf{A}(\Omega,\mathcal{W})=\left.\left(\begin{array}{cc}
        \textbf{0} &  \textbf{I}\\
        - \textbf{M}^{-1}\textbf{K} & -\textbf{M}^{-1}\textbf{C}_{NL}
    \end{array}\right)\right|_{\mathbf{x} = \mathbf{x}^*}\,,
\end{equation}

\noindent where  $\textbf{C}_{NL}$ differs from $\textbf{C}$ only by the last component $({n_{el}},{n_{el}})$ which must add the derivative related to the nonlinear bit-rock interaction model, $\partial T_{bit}/\partial \dot{\theta}_{n_{el}}$ (see Section \ref{sec:stability}), $\textbf{I}$ and $\textbf{0} \in \mathbb{R}^{n_{el}\times n_{el}}$ are the identity and the zero matrices.  
If the real part of the eigenvalues of $\textbf{A}$ are all negative the system is stable.

\begin{figure}[H]
\begin{center}
\includegraphics[scale=.35]{figures/MapDet.eps}
\includegraphics[scale=.35]{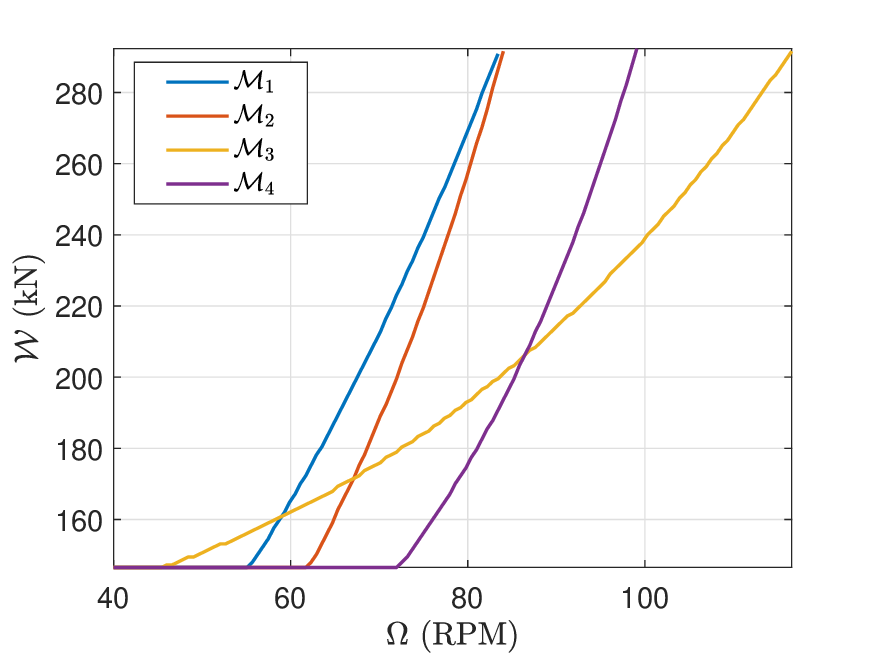}
\caption{Left: stability map for the 1-DOF system. Right: stability map for the FE model 2-DOF (one element for the drill-pipe and one for the BHA). Operational points ($\Omega,\mathcal{W}$) in the region above the curves are unstable and below are stable.}
\label{fig:boundary_modelsID2_2}
\end{center}
\end{figure}

The values of the parameters used in the simulation are shown in Table \ref{tb:drill_string_geometry} \cite{Ritto2017}; outer and inner diameters are given in the table. Figure \ref{fig:boundary_modelsID2_2} shows how similar the stability maps are for the 1-DOF and the 2-DOF FE systems. The 1-DOF system  has $\omega_n=0.85$ rad/ and $\xi=0.25$, and the 2-DOF FE system has $\omega_n=\{0.85, 15.60\}$ rad/s and $\xi=\{0.30,0.06\}$, where the damping parameters are $\alpha=0.5$ and $\beta= 0.006$. 

\begin{figure}[H]
\begin{center}
\includegraphics[scale=.35]{figures/MapEF2DOF.eps}
\includegraphics[scale=.35]{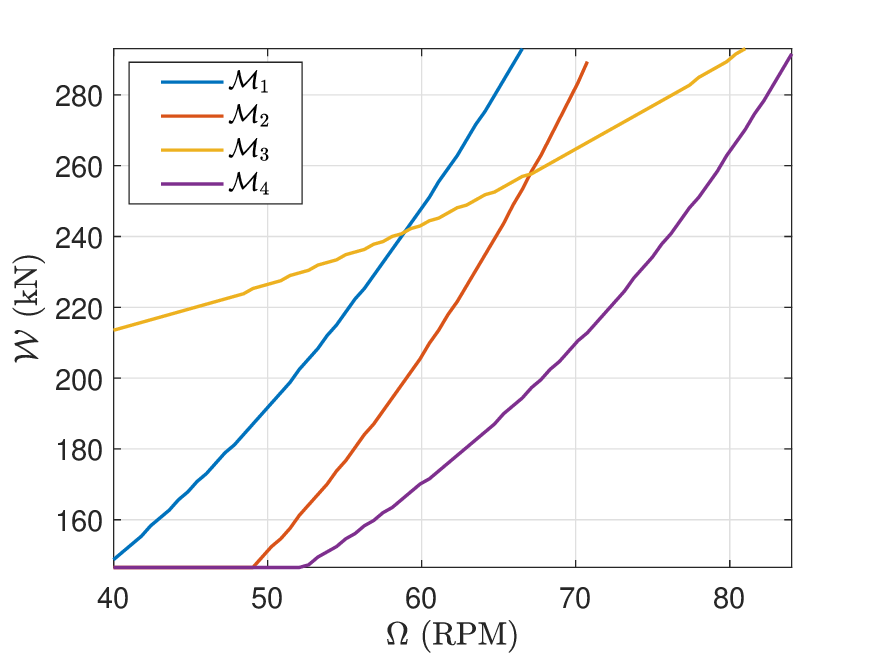}
\caption{Left: stability map for FE model 2-DOF (one element for the drill-pipe and one for the BHA). Right: stability map for FE model 10-DOF (eight elements for the drill-pipe and two for the BHA); $\alpha=0.5$ and $\beta= 0.006$. Operational points ($\Omega,\mathcal{W}$) in the region above the curves are unstable and below are stable.}
\label{fig:FE2DOF}
\end{center}
\end{figure}

Figure \ref{fig:FE2DOF}
shows the stability map comparing the 2-DOF with the 10-DOF FE discretization using the same proportional constants $\alpha=0.5$ and $\beta= 0.006$. It is notorious that the stability boundary drawn by the models changes considerably. For the 10-DOF system
$\omega_n=\{0.83, 2.66, 4.76, 7.11,9.73, 12.62,15.63,$ $18.23, 22.75, 45.00\}$ rad/s and  $\xi=\{0.30, 0.10, 0.07,0.06,0.05,0.06,0.06,0.07, 0.08,$ $0.14\}$. However, we argue that this might be an unfair comparison because one could adjust the damping parameters to get a similar stability boundary, and it is well known that damping plays a significant role in stability.

\begin{figure}[H]
\begin{center}
\includegraphics[scale=.35]{figures/MapEF2DOF.eps}
\includegraphics[scale=.35]{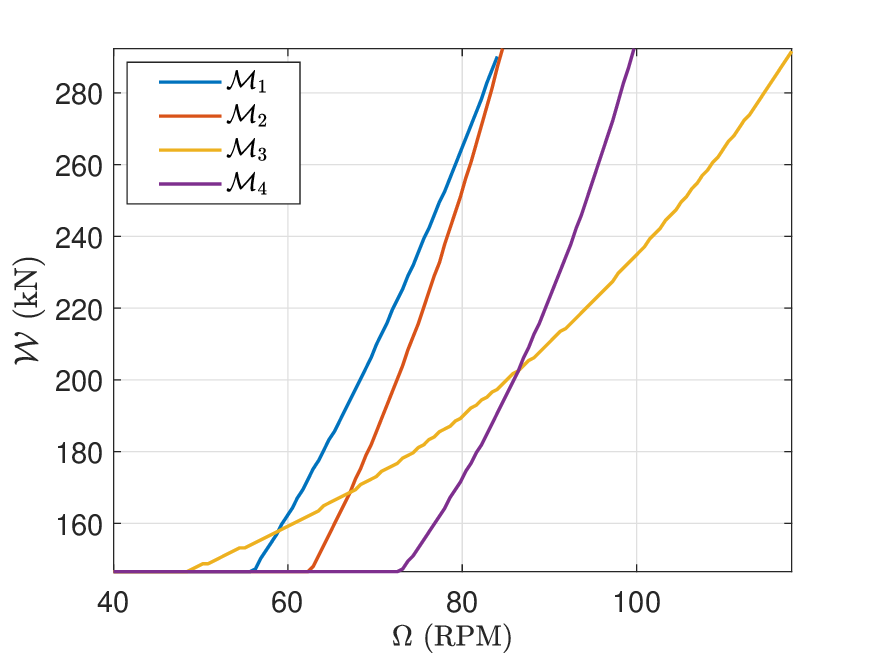}
\caption{Left: stability map for FE model 2-DOF (one element for the drill-pipe and one for the BHA); $\alpha=0.5$ and $\beta= 0.006$. Right: stability map for FE model 10-DOF (eight elements for the drill-pipe and two for the BHA); $\alpha=0.5$ and $\beta= 0.0021$. Operational points ($\Omega,\mathcal{W}$) in the region above the curves are unstable and below are stable.}
\label{fig:FE10DOF_2}
\end{center}
\end{figure}

Figure \ref{fig:FE10DOF_2} shows the stability map comparing the 2-DOF with the 10-DOF FE discretization using a slightly smaller damping values for the 10-DOF system, $\alpha=0.5$ and $\beta= 0.0021$. Now for the 10-DOF system, the natural frequencies remain the same and the damping rates are slightly reduced $\xi=\{0.30, 0.10, 0.06, 0.04, 0.03, 0.03, 0.03, 0.03, 0.03, 0.05\}$. The figure shows that the stability maps are now equivalent.

The conclusion is that if the number of elements increases a different map is obtained, but if the damping parameters are adjusted we get equivalent results. The damping affects the system stabilizing it, and its modeling is always a challenge. The simple proportional damping model is useful in many situations, including the drill-string problem, but its coefficients must be identified for the specific defined FE discretization.
	
Finally, it should be highlighted that the FE drill-string model may be necessary when higher modes are activated. This might happen when analyzing other types of instabilities, coupling mechanisms, impact motion, friction and taping along the column, stochastic perturbations, etc.
	
\end{document}